\documentclass[fleqn,usenatbib]{mnras}

\usepackage{newtxtext,newtxmath}

\usepackage[T1]{fontenc}
\usepackage{ae,aecompl}

\usepackage{bm}
\let\vec\bm

\usepackage{mleftright}
\let\left\mleft
\let\right\mright

\usepackage{graphicx}
\graphicspath{{figures/}}
\usepackage{amsmath}

\usepackage[normalem]{ulem}

\newcommand{\bs}[1]{\boldsymbol{#1}}
\newcommand{\be}{\begin{equation}}
\newcommand{\en}{\end{equation}}

\newcommand{\pder}[2]{\frac{\partial{#1}}{\partial{#2}}}
\newcommand{\pdder}[2]{\frac{\partial^2{#1}}{\partial{#2}^2}}

\newcommand{\der}[2]{\frac{\mathrm{d}#1}{\mathrm{d}#2}}

\newcommand{\f}{\frac}

\newcommand{\ex}{\vec{e}_x}
\newcommand{\ey}{\vec{e}_y}
\newcommand{\ez}{\vec{e}_z}

\newcommand{\para}{\parallel}

\newcommand{\paren}[1]{ \left( #1 \right) }
\newcommand{\kb}{k_{\textup {B}}}

\renewcommand{\b}{\bb{b}}

\newcommand{\dvx}{\delta \varv_{x}}
\newcommand{\dvz}{\delta \varv_{z}}

\newcommand{\dt}{\frac{\delta T}{T}}
\newcommand{\drho}{\frac{\delta \rho}{\rho}}
\newcommand{\dA}{\f{\delta A}{B}}

\newcommand{\mH}{m_{\textup{H}}}

\newcommand\bb[1]{\mbox{\boldmath{$#1$}}}
\newcommand\bcdot{\bb{\cdot}}
\newcommand\del{\bs{\nabla}}

\newcommand{\va}{\varv_\mathrm{a}}

\newcommand{\cs}{c}

\newcommand{\ui}{\mathrm{i}}

\newcommand{\Kni}{\mathrm{Kn}_0^{-1}}
\newcommand{\Kn}{\mathrm{Kn}_0}
\newcommand{\Sd}{S_\mathrm{D}}
\newcommand{\omdy}{\omega_{ \mathrm{dyn}}}
\newcommand{\omc}{\omega_{ \mathrm{c}}}

\newcommand{\oma}{\omega_{\mathrm{a}}}

\newcommand{\HO}{H}

\title[Suppressed heat conductivity in the ICM]{Suppressed heat conductivity
in the intracluster medium: implications for the magneto-thermal instability}

\author[Berlok, Quataert, Pessah \& Pfrommer]{
Thomas Berlok$^{1,2}$\thanks{E-mail: tberlok@aip.de}, Eliot Quataert$^{3,4}$,
Martin E. Pessah$^{2}$
and Christoph Pfrommer$^{1}$
\\
$^{1}$Leibniz-Institut f{\"u}r Astrophysik Potsdam (AIP), An der Sternwarte 16, D-14482
 Potsdam, Germany \\
$^{2}$Niels Bohr International Academy, Niels Bohr Institute, Blegdamsvej 17, DK-2100
 Copenhagen {{\O}}, Denmark \\
$^{3}$Department of Astrophysical Sciences, Princeton University, Princeton, NJ 08544, USA \\
$^{4}$Department of Astronomy and Theoretical Astrophysics Center, University
of California, Berkeley, CA 94720-3411, USA
}

\date{Accepted XXX. Received YYY; in original form ZZZ}

\pubyear{2020}

\usepackage[dvipsnames]{xcolor}

\begin{document}
\label{firstpage}
\pagerange{\pageref{firstpage}--\pageref{lastpage}}
\maketitle

\begin{abstract}
In the outskirts of the intracluster medium (ICM) in galaxy clusters,
 the temperature   decreases with radius.
Due to the weakly collisional nature of the plasma, these regions
 are susceptible to the magneto-thermal instability (MTI), which can sustain
turbulence and provide turbulent pressure support in the ICM.
This instability arises due
to heat conduction directed along the magnetic field, with a heat conductivity
which is normally assumed to be given by the Spitzer value. Recent numerical
studies of the ion mirror and the electron whistler instability
using particle-in-cell codes have  shown that microscale instabilities
can lead to a reduced value for the heat conductivity in the ICM. This
could in turn influence the efficiency with which the MTI drives turbulence. In
this paper we investigate the influence of reduced heat transport on the
nonlinear evolution of the MTI. We study plane-parallel, initially
static atmospheres and  employ a subgrid model that mimics the
influence of the mirror instability
on the heat conductivity. We use this subgrid model to assess the effect of
microscales on the large scale dynamics of the  ICM. We find
that the
nonlinear saturation of the
MTI is surprisingly
robust in our simulations.
Over a factor of $\sim 10^3$ in the thermal-to-magnetic pressure ratio and
collisionality we find at most modest changes to the saturation of the MTI
with respect to reference simulations where heat transport is unsuppressed.
  \end{abstract}


\section{Introduction}

The intracluster medium of galaxy clusters consists of a plasma with high
temperature ($T\sim 10^7$--$10^8 $~K) and low number density
($n\sim 10^{-4}$--$10^{-2}~\rmn{cm}^{-3}$). This plasma is weakly collisional
and has, as a result, convective stability properties that differ from those expected
 from
the Schwarzschild criterion \citep{1958ses..book.....S}. The new stability
properties arise as a consequence of anisotropic transport of heat, in which
the heat flow is directed along the local magnetic field. The physics of this
anisotropic transport has been modelled using Braginskii MHD \citep{Bra}
in which the heat flow vector, $\vec{Q}$, is given by
\be
    \vec{Q} = -\chi_\para \b\b \cdot \nabla T,
    \label{eq:heat-flux}
\en
where $\chi_\para$ is the heat conductivity
\citep{Spitzer1962}, $\b$ is a unit vector directed
along the magnetic field, $\b\b$ is a dyadic product, and
$T$ is the temperature of the plasma.

Two of the widely studied instabilities that arise due to anisotropic
transport of heat are the magneto-thermal instability (MTI,
\citealt{Bal00,Bal01}) and the heat-flux-driven buoyancy instability (HBI,
\citealt{Qua08}). The MTI  can be active for stratified atmospheres in which
the temperature decreases with height and has been studied using Braginskii
MHD \citep{Par05,Par08_MTI,Kun11,Kun12,2012MNRAS.419L..29P,Par12}.

While Braginskii MHD can describe some features of weakly collisional systems
it does not correctly describe all the microscale instabilities that a weakly
collisional plasma such as the ICM can support
\citep{schekochihin_plasma_2005,schekochihin_fast_2006}. In fact, fully
collisionless plasmas are better described by the Vlasov-Maxwell equations
\citep{Ichimaru1973,Krall1973,Swanson1989,Stix1992}. This system of equations
can be used to derive the linear theory of kinetic instabilities such as the
mirror and firehose instabilities
\citep{Chandrasekhar1958,Parker1958,Hasegawa1969,Gary1993} which are driven by
anisotropy in velocity space with respect to the magnetic field. The nonlinear
evolution of these microscale instabilities can be studied by employing
particle-in-cell simulations \citep[PIC,][]{Hockney1988,Birdsall1991}.

 Studies of velocity-space instabilities in collisionless plasmas have
shown that the effective heat conductivity can be reduced by the ion mirror
instability \citep{Kom16,Riq16} and the electron whistler instability
\citep{Roberg-Clark2016ApJ,Riq16,Roberg-Clark2018PhRvL,Roberg-Clark2018ApJ,Komarov2018}.
 This suppression of heat conductivity could potentially
influence the evolution and saturation of the MTI and the HBI which both have
growth rates that depend on the magnitude of the anisotropic heat
conductivity, $\chi_\para$. We focus in the following on the MTI because it is
expected to be active in the outskirts of clusters where the plasma-$\beta$ is
higher and the collisionality is lower than in the inner regions of the
cluster. Both a weak magnetic field and a low collisionality conspire to make
a plasma prone to the mirror instability and it seems likely that their
presence can influence the dynamics of the ICM in this region.

While PIC simulations of the Vlasov-Maxwell systems can be used to study
small-scale instabilities in the ICM they are so far too computationally expensive
to be used for studies of the large scale dynamics. In this
paper we instead attempt to bridge the two approaches (PIC and Braginskii-MHD)
by using published PIC simulations to motivate a subgrid model for the heat
conductivity in the ICM. We use this subgrid model for the heat conductivity
in simulations of the MTI in order to assess whether the
microscale instabilities modify its evolution.

Our study is performed using a plane-parallel approximation to the ICM
atmosphere which we assume to be initially in hydrostatic equilibrium.
We are thus neglecting the effects of turbulence driven by e.g. AGN or mergers
\citep{Zuhone2011,McNamara2012,Kannan2017,Barnes2019}.
Such external turbulence can modify the MTI directly by randomizing field lines \citep{2011MNRAS.413.1295M,Ruszkowski2011ApJ}
and/or indirectly by increasing the volume fraction of the ICM which is mirror unstable
(e.g. \citealt{Santos-Lima2014ApJ}) and has a lower heat conductivity.
This interaction between AGN /merger driven turbulence
and MTI depends on the details of the turbulence (e.g. its strength and injection scale
relative to the atmosphere scale height, see \citealt{2011MNRAS.413.1295M})
and is likely to have a time-dependent character since the AGN/merger driving changes
in time. In this paper we only provide a qualitative discussion
of external turbulence as a detailed understanding of this interaction requires dedicated
cosmological simulations.

The rest of the paper is outlined as follows. In
Section~\ref{sec:methods}, we introduce the equations of Braginskii-MHD along
with subgrid models for viscosity (Section~\ref{sec:visc-subgrid}) and heat
conduction (Section~\ref{sec:heat-subgrid}). In Section~\ref{sec:MTI}, we then
study how the MTI is modified by suppression of heat conduction in mirror
unstable regions. We present our numerical setup
(Section~\ref{sec:numerical_setup}), expectations from local linear theory
(Section~\ref{sec:mti-local-theory}) and a simulation of the interruption of a
single eigenmode of the MTI (Section~\ref{sec:mti-interrup}). We analyze a 2D
nonlinear MTI simulation (Section~\ref{sec:nonlinear}), perform a parameter
study (Section~\ref{sec:parameter_study}), and compare our 2D findings with 3D
simulations (Section~\ref{sec:3D}). We summarize and discuss our results and its limitations (e.g. the role of external turbulence and other plasma instabilities) in
Section~\ref{sec:discussion}. The appendices include additional details on
how sound waves are modified by our subgrid model for heat conductivity
(Appendix~\ref{sec:soundwave}), the quasi-global linear theory for the MTI
(Appendix~\ref{app:quasi-global-theory}) and a resolution study of our
simulations (Appendix~\ref{app:resolution-study}).
\section{Fluid description of weakly collisional plasmas - equations
and subgrid models}
\label{sec:methods}

\subsection{Equations of Braginskii-MHD}
\label{sec:equations}

In Braginskii-MHD, the mass continuity equation, the momentum equation, the
induction equation, and the energy equation are given by
\citep{Bra,1983bpp..conf....1K,Sch10}
\be
    \pder{\rho}{t} + \nabla \cdot(\rho \vec{\varv}) = 0 \ ,
    \label{eq:rho}
\en
\be
    \pder{\paren{\rho \vec{\varv}}}{t}
+\nabla \cdot \paren{\rho \vec{\varv \varv} + p_{\mathrm{T}} \mathbf{1}
- \frac{B^2}{4\upi} \b\b}
=
-\nabla \cdot \vec{\Pi}
+ \rho \vec{g}  \ ,
\en
\be
    \pder{\vec{B}}{t} = \nabla\times(\vec{\varv}\times\vec{B}) \ ,
    \label{eq:b}
\en
\be
    \pder{E}{t} + \nabla \cdot \left[ \paren{E+p_{\mathrm{T}}}\vec{\varv} -
    \frac{\vec{B}\paren{\vec{B\cdot \varv}}}{4\upi}\right]
    = -\nabla \cdot \vec{Q}
      - \nabla \cdot
    \paren{\vec{\Pi} \cdot \vec{\varv}} +\rho \vec{g\cdot \varv} \ ,
    \label{eq:energy}
\en
where $\rho$ is the mass density, $\vec{\varv}$ is the fluid velocity,
$\vec{B}$ is the magnetic field with direction $\b$ and $p_{\mathrm{T}}=p +
B^2/8\upi$ is the total (gas plus magnetic) pressure. Here the total energy
density is
\be
    E = \f{1}{2}\rho \varv^2 + \f{B^2}{8\upi} + \f{p}{\gamma - 1} \ ,
\en
with $\gamma = 5/3$ and $p = \rho \kb T/\mH \mu$ where $\kb$ is Boltzmann's
constant, $\mH$ is the proton mass and $\mu$ is the mean molecular weight.
Gravity is included in the momentum and energy equations.  In our
numerical models we will take the
gravitational acceleration  to be  $\vec{g} = -g \ez$
 where $g$ is a constant.

The equations of Braginskii-MHD, equations~\eqref{eq:rho}-\eqref{eq:energy},
differ from ideal MHD by the inclusion of two diffusive effects, anisotropic
heat conduction and viscosity (the latter known as Braginskii viscosity).
The anisotropic heat flux, $\vec{Q}$, is present in the energy equation,
equation~\eqref{eq:energy}, and has already been introduced in
equation~\eqref{eq:heat-flux}. It describes electron heat conduction which
is directed along the direction of the magnetic field. Its magnitude depends
on the heat conductivity, $\chi_\para$, which is related to
the heat diffusivity by $\kappa_\para = \chi_\para T/p$.

The anisotropic viscosity tensor, $\vec{\Pi}$, enters in both the momentum
and energy equations, and is given by
\be
    \vec{\Pi} = -\Delta p \left(\b\b -\f{1}{3} \mathbf{1}\right) \ ,
\en
where $\Delta p = p_\perp - p_\para$ is the pressure anisotropy and $p_\para$
($p_\perp$) is the pressure parallel (perpendicular) to the magnetic field. In
a weakly collisional plasma, the pressure anisotropy can be assumed to be
given by \citep{schekochihin_plasma_2005}
\be
    \label{eq:pressure_anisotropy}
    \Delta p = \rho \nu_\para
     \der{\ln B^3 \rho^{-2}}{t} =
    \rho \nu_\para \paren{3 \b\b \vec{:} \nabla \vec{\varv} -
          \nabla \cdot \vec{\varv}}
    \ ,
\en
where $\nu_\para$ is the viscosity coefficient and
$\mathrm{d}/\mathrm{d}t=\partial /\partial t + \vec{\varv} \bcdot \del $
is the Lagrangian derivative.

\subsection{Transport coefficients}
\label{sec:coeffs}

The values of $\nu_\para$ and $\chi_\para$ are set by the collisionality of
the plasma. In standard Braginskii-MHD, these are
\be
    \label{eq:chi_aniso}
    \chi_\para = \f{5 p_\mathrm{e}}{2 m_\mathrm{e}\nu_\mathrm{ee}} \ , \\
    \nu_{\para} = \f{p_\mathrm{i}}{\rho\nu_\mathrm{ii}} \ ,
\en
where $p_\mathrm{i}$ ($p_\mathrm{e}$) is the ion (electron) pressure,
$\nu_\mathrm{ii}$ ($\nu_\mathrm{ee}$) is the ion-ion (electron-electron)
Coulomb collision frequency and $m_\mathrm{e}$ is the electron mass.

The Knudsen number, $\mathrm{Kn}$, is a useful measure of the
collisionality of the plasma with the collisionless (collisional) limit
corresponding to $\mathrm{Kn}^{-1} \ll 1$ ($\mathrm{Kn}^{-1} \gg 1$).
It is defined as
\be
    \mathrm{Kn} \equiv \f{\lambda_\mathrm{i}}{H} \ ,
\en
where $\lambda_\mathrm{i}=\cs/\nu_\mathrm{ii}$ is the ion mean free path,
 $\cs =
\sqrt{p/\rho}$ is the isothermal sound speed, and $H=\cs^2/g$ is the pressure
scale height of the plasma.
In terms of the Knudsen number, the heat diffusivity, $\kappa_\para =
\chi_\para T/p$, and the Braginskii viscosity coefficient, $\nu_\para$, are
given by \citep{Kun12}
\be
    \label{eq:kappa_and_nu_para}
    \kappa_\para \approx 24\,\mathrm{Kn}\, \cs H \ , \\
    \nu_\para \approx 0.48\,\mathrm{Kn}\, \cs H \ .
\en
Note here that the collisionless limit,
$\mathrm{Kn}^{-1} \ll 1$, corresponds to high values of
$\kappa_\para$ and $\nu_\para$, i.e., fast transport. Finally, it is useful
to define the plasma-$\beta$ as $\beta = {2
\cs^2}/{\va^2}$ where the Alfv\'{e}n speed is given by $\va =
{B}/{\sqrt{4\upi \rho}}$.
In the outskirts of galaxy clusters, estimated values for the
dimensionless numbers are
$\beta\sim 10^2$--$10^4$
and $\mathrm{Kn}^{-1}\sim10$--$100$
\citep{Car02,Vik06}.

\subsection{The 'standard' subgrid model for viscosity}
\label{sec:visc-subgrid}

Equation~\eqref{eq:pressure_anisotropy} shows that a pressure anisotropy
naturally arises in a weakly collisional plasma because of changes in magnetic
field strength or plasma density. The mirror and firehose instabilities become
active if the magnitude of the pressure anisotropy is comparable to the
magnetic energy density. This is more likely to occur if the magnetic field
strength is weak, as it is in the ICM. These microscale instabilities are not
correctly described by the equations of Braginskii MHD. PIC simulations (e.g.
\citealt{Kunz2014a}) and solar wind observations (e.g.
\citealt{Bale2009,Chen2016}) indicate that the microscale instabilities grow
extremely quickly and act to remove excess pressure anisotropy to sustain
marginal stability on average. Motivated by these studies, a common subgrid
model for $\Delta p$ consists of limiting its value to lie within the
thresholds for stability of the firehose and mirror instability (e.g.
\citealt{Sha06,Kun12,Squire2017}). We employ this subgrid model and limit
$\Delta p$ by
\be
    \label{eq:fire_and_mirror}
    -\f{B^2}{4\upi} < \Delta p < \f{B^2}{8\upi} \ ,
\en
when evaluating the viscosity tensor in our simulations.

\subsection{The subgrid model for suppressed heat conduction}
\label{sec:heat-subgrid}

\citet{Kom16} found that the heat conductivity in a
mirror-unstable plasma is reduced by a suppression
factor $S_\mathrm{D} \approx 0.2$ compared to the Spitzer value.
This result was found by extracting magnetic field
lines from the hybrid-kinetic (kinetic ions, fluid electrons) PIC simulation
presented in \citet{Kunz2014a} and studying the effect of magnetic mirrors on
the electron dynamics. The suppression of heat conductivity arises due to a
combination of electron trapping in magnetic mirrors (which prevents them from
contributing to conduction) and a decrease in the effective mean-free-path of
collisions for the untrapped electrons.

Motivated by these results, we replace $\chi_\para$ with an effective heat
conductivity
\be
    \chi_{\textrm{eff}} = S_\mathrm{D}\chi_\para
\en
in the regions where the ions are unstable according to the linear
stability criterion for the mirror instability. This subgrid model for the
heat conductivity is designed to mimic the behaviour seen in more realistic
(but small-scale) models in the simplest manner. The mirror instability
becomes unstable when
\be
  \Delta p \gtrapprox \f{B^2}{8\upi} \ ,
  \label{eq:mirror}
\en
and by combining equation~\eqref{eq:pressure_anisotropy} and
equation~\eqref{eq:kappa_and_nu_para} we can rewrite this condition for the
mirror instability (and the consequent suppression of heat conductivity) can be rewritten as
\be
  0.48
    \paren{3 \b\b \vec{:} \nabla \vec{\varv} -
          \nabla \cdot \vec{\varv}} \f{H}{\cs}
  \gtrapprox
  \f{\mathrm{Kn}^{-1}}{\beta} \ .
  \label{eq:mirror-Knbeta}
\en
Equation~\eqref{eq:mirror-Knbeta} predicts that plasmas are more prone to the
mirror instability when $\beta \mathrm{Kn} \gg 1$, i.e., when the magnetic
field is weak and collisions are rare. The subgrid model for suppression of
heat conductivity is thus predicted to become active in this limit.
We find it is useful to understand how plasma dynamics are modified in
the two extreme regimes, $\Sd=0.01$ (almost full suppression) and $\Sd=1$
(no suppression), and consequently perform most of our simulations with
these parameters. We do however return to the value motivated by \citet{Kom16},
$S_\mathrm{D} = 0.2$, in Section~\ref{sec:discussion}.

\section{The magneto-thermal instability}
\label{sec:MTI}

\subsection{Numerical setup}
\label{sec:numerical_setup}

We consider the numerical setup described in \cite{Kun12}, i.e. an MTI
unstable region with anisotropic heat conductivity and Braginskii viscosity
sandwiched between two stable regions with isotropic heat conductivity (see
also \citealt{Par07}). For completeness, the profiles for the
temperature\footnote{We assume that the ions and electrons have the same
temperature. The low collisionality however leads to a long
temperature
equilibration time scale in the outskirts of clusters.}, $T(z)$, and density,
$\rho(z)$, are given by \citep{Par05,2011MNRAS.413.1295M,Kun12}
\be
    \rho(z) = \rho_0 \left(1 - \f{z}{3\HO}\right)^2 \ ,
    \label{eq:density-profile}
\en
\be
    \label{eq:temperature-profile}
    T(z) = T_0 \left(1 - \f{z}{3\HO}\right) \ ,
\en
where $z$ is the vertical coordinate and $\HO$ is the scale height at the
lower boundary of the unstable domain. These profiles satisfy hydrostatic
equilibrium
\be
    \pder{p}{z} = -\rho g \ ,
    \label{eq:pressure-profile}
\en
in a (constant) gravitational field with $g=k_\rmn{B} T_0 / (\HO\mu \mH)$.
Here the pressure is given by the
equation of state in which we assume that the composition of the plasma is
constant.\footnote{See \cite{Pes13}, \cite{Berlok2015}, \cite{Berlok2016a},
and \cite{Berlok2016b} for studies that relax this assumption.} The
temperature decreases in the direction opposite to the direction of gravity.
We include an initially horizontal magnetic field which makes the atmosphere
maximally unstable to the MTI. The magnetic field strength is spatially
constant and is characterized by the plasma-$\beta$ at the bottom of the
unstable domain, $\beta_0$. In our simulation analysis, the energy
densities are given in units of $\rho_0 c_0^2$ where $\rho_0$ ($c_0$) is the
density (isothermal sound speed) at the bottom of the unstable domain.
Similarly, the inverse Knudsen number at this location is designated $\Kni$.
The boundary conditions at the top and bottom of the computational
domain maintain hydrostatic equilibrium by extrapolating density and
pressure while keeping the temperature fixed. In addition, the velocity is
reflected and the magnetic field is forced to remain horizontal.
Our boundary conditions are described in further detail in Appendix B1 in
\citet{Berlok2016a}.

We use the publicly available MHD code Athena
\citep{stone_athena:_2008,stone_simple_2009} to evolve
equations~\eqref{eq:rho}-\eqref{eq:energy} in time. The implementation of
anisotropic heat conduction is described in \cite{Par05,Sha07} and anisotropic
viscosity in \cite{Par12}. The time step constraint on explicit solution of
these parabolic diffusion equations makes the simulations very computationally
demanding. We use sub-cycling with 10 diffusion steps per MHD step as in
\cite{Kun12} but note that future studies could likely benefit from using
super timestepping methods \citep{Meyer2012,Vaidya2017,Berlok2019c}.
We present two-dimensional (2D) simulations
in Sections~\ref{sec:mti-interrup}--\ref{sec:parameter_study}
and three-dimensional (3D) simulations in Section~\ref{sec:3D}.
Tables~\ref{tab:simulations} and
\ref{tab:convergence-simulations} in Appendix~\ref{app:resolution-study}
give an overview of the parameters and numerical resolutions used in the
simulations.

\subsection{Expectations from local, linear theory}
\label{sec:mti-local-theory}

The MTI arises as a consequence of heat conduction along magnetic field lines
in a thermally stratified atmosphere \citep{Bal00,Bal01}. The growth rate of
the MTI is therefore intimately connected to the magnitude of the heat
conductivity, $\chi_\para$. Here we use the local, linear theory of
\citet{Kun11} to estimate the change in growth rate when the heat conductivity
is suppressed by a factor $\Sd=0.01$.

\begin{figure}
    \centering
    \includegraphics[trim = 0 20 0 0]{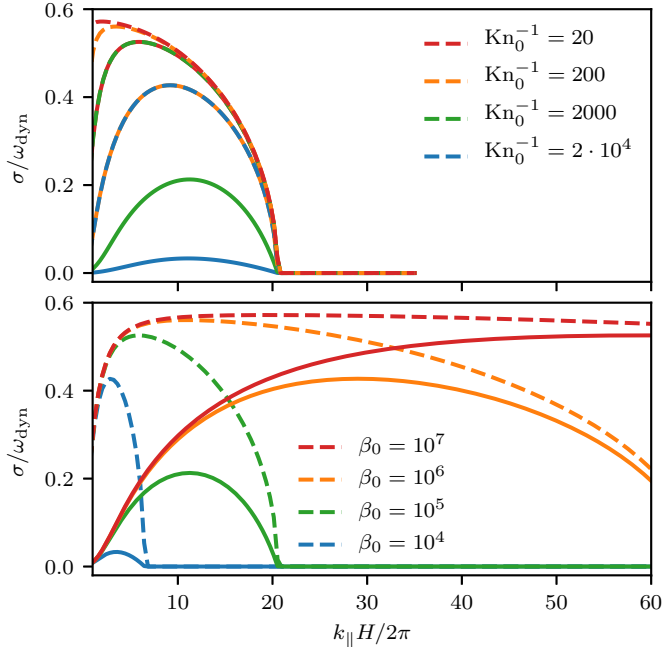}
 \caption{Growth rates as function of mode number, $n=k_\para H/2\upi$, calculated at the
    bottom of the unstable domain, for a variety of collisionalities for
    $\beta_0=10^5$ (upper panel) and for a variety of $\beta_0$'s for $\Kni=
    2000$ (lower panel). Solid lines show theoretical growth rate for
    a heat conductivity which is smaller by a factor $S_\mathrm{D}=0.01$
    with respect to the corresponding  dashed line.}
\label{fig:theoretical_growth_rates}
\end{figure}

\citet{Kun11} included the effect of Braginskii viscosity in the derivation of
a local, linear dispersion relation for the MTI. It was found that the maximum
growth rate occurs for wave vectors parallel to the magnetic field, $k=k_x$,
and that this mode of instability is unaffected by Braginskii viscosity. We
therefore calculate the MTI growth rate as a function of $k=k_x$ for various
values of $\Kni$ (which sets the heat conductivity, see
equation~\ref{eq:kappa_and_nu_para}) and $\beta_0$.
The results are shown in Fig.~\ref{fig:theoretical_growth_rates} where $\Kni$ is
varied in the upper panel at fixed $\beta_0=10^5$ and $\beta_0$ is varied in
the lower panel at fixed $\Kni=2000$. Here the growth rate, $\sigma$,
is measured with respect to the dynamical frequency, $\omdy = c_0/\HO$, at
the bottom of the unstable domain.
The upper panel of Fig.~\ref{fig:theoretical_growth_rates} shows that
increasing $\Kni$ decreases the growth rate at all wavelengths. At the same
time, the fastest growing mode is shifted to shorter wavelengths. As
previously explained, this was expected as the MTI depends on efficient heat
conduction and higher $\Kni$ translates to lower $\chi_\para$. Each
dashed line has a corresponding solid line (with the same color) where the heat
 conductivity has been suppressed by
a factor $\Sd=0.01$. Interestingly,  the reduction in
MTI growth
rate  depends on how fast the heat conduction is without suppression. For
instance, the maximal growth rate for $\Kni=2\times10^{4}$ (blue
 dashed curve) is $\sigma_{\mathrm{max}}/\omdy=0.43$ while it
is only $\sigma_{\mathrm{max}}/\omdy=0.03$ when suppressed (blue
 solid curve). With faster heat conduction, e.g. $\Kni=20$,
the fastest growth rate is only slightly reduced (from
$\sigma_{\mathrm{max}}/\omdy=0.57$ to $\sigma_{\mathrm{max}}/\omdy=0.53$). In
conclusion, suppression of heat conduction completely quenches the MTI if the
heat conductivity is already low but if the unsuppressed heat conductivity is
high, the suppression of heat conduction only leads to a moderate reduction in
the growth rate.

In the lower panel of Fig.~\ref{fig:theoretical_growth_rates} we show the
dependence of the growth rates on $\beta_0$. The trend is that the growth
rates increase with $\beta_0$ and that the fastest growing mode shifts to
shorter wavelengths. The explanation for this behavior is that magnetic
tension inhibits or even quenches the instability when $\oma \gg \omdy$ where
$\oma = k_\para \va$ is the Alfv\'{e}n frequency.. This means that growth of the MTI is
 prevented when
$k_\para H \gg \sqrt{\beta_0/2}$. Each  dashed line again has
a corresponding  solid line where the heat conductivity is
suppressed with $\Sd=0.01$. At low $\beta_0$ ($10^4$ and $10^5$), the
suppression of heat conductivity leads to a severe reduction in the MTI growth
rate. At high $\beta_0$ ($10^6$ and $10^7$), the reduction in the maximal
growth rate is however only moderate.

The  results in
Fig.~\ref{fig:theoretical_growth_rates} can be explained in terms of the ratio
of
conduction frequency to the dynamical frequency. Here the conduction frequency
\be
    \label{eq:omega_c}
    \omc = \f{2}{5} \,\kappa_\para k_\para^2 \ ,
\en
sets the inverse time scale for heat conduction across a mode with parallel
wavelength, $k_\para$. Significant growth of a given MTI mode depends on
whether it is in the fast conduction limit, $\omc/\omdy > 1$ or not. Whether a
mode is quenched by suppression of heat conductivity (or whether the growth
rate is only moderately modified) then depends on whether the mode retains
$\omc/\omdy > 1$ or whether $\omc/\omdy < 1$ after suppression.

As a function of the mode number, $n = k_\para H/2\upi$ and using
equation~\eqref{eq:kappa_and_nu_para} the ratio $\omc/\omdy$ can be estimated
as
\be
    \f{\omc}{\omdy}
    \approx  S_\mathrm{D} \f{380}{\Kni} \, n^2 \ ,
    \label{eq:fast-cond-limit}
\en
At high $\beta_0$, $\omc/\omdy$ remains large because the fastest growing mode
has very high $n$ (due to  weak magnetic tension). Physically, the
time scale for heat transport across the mode is short even with suppression
because the spatial scales are small.

More quantitatively, magnetic tension prevents growth when $k_\para H \gg
\sqrt{\beta_0/2}$, such that the mode number that gives the maximum growth
rate, $n_\mathrm{max}$, will scale as $\sqrt{\beta_0}$. This leads to the
scaling $\omc/\omdy \propto \Sd \beta_0/\Kni$. The linear theory thus indicates that the
 fastest growing MTI
mode can remain in the fast heat conduction limit if $\beta_0/\Kni\gg1$ is
large. Incidentally, $\beta_0/\Kni\gg1$ large is precisely the limit where the
mirror instability is likely to take place and lead to suppression of heat
conductivity (see equation~\ref{eq:mirror-Knbeta}). In the opposite limit,
$\beta_0/\Kni$ small, the MTI growth rate can be severely reduced by a
suppression of heat conductivity. In this limit, however, such suppression is
unlikely to take place because the mirror instability is not as easily excited
(see also equation~\ref{eq:mirror-Knbeta}).

In conclusion, both the MTI and the mirror instability take their most
vigorous form in a plasma in which the magnetic field is weak (large
$\beta_0$) and the collisionality is low (small $\Kni$). In this regime, the
MTI grows fast because \emph{i)} the heat conductivity, $\chi_\para$, is high
and \emph{ii)} a weak magnetic field only prevents growth on small scales
(across which heat conduction remains fast). In this regime, the MTI growth
rates remain dynamically important even if the heat conductivity is suppressed
by a significant factor, e.g., $\Sd=0.01$. This observation will become
important for interpreting our nonlinear simulations in
Section~\ref{sec:parameter_study}.

\subsection{Interruption of an eigenmode of the MTI}
\label{sec:mti-interrup}

\begin{figure*}
\centering
\includegraphics[trim= 15 40 0 60]{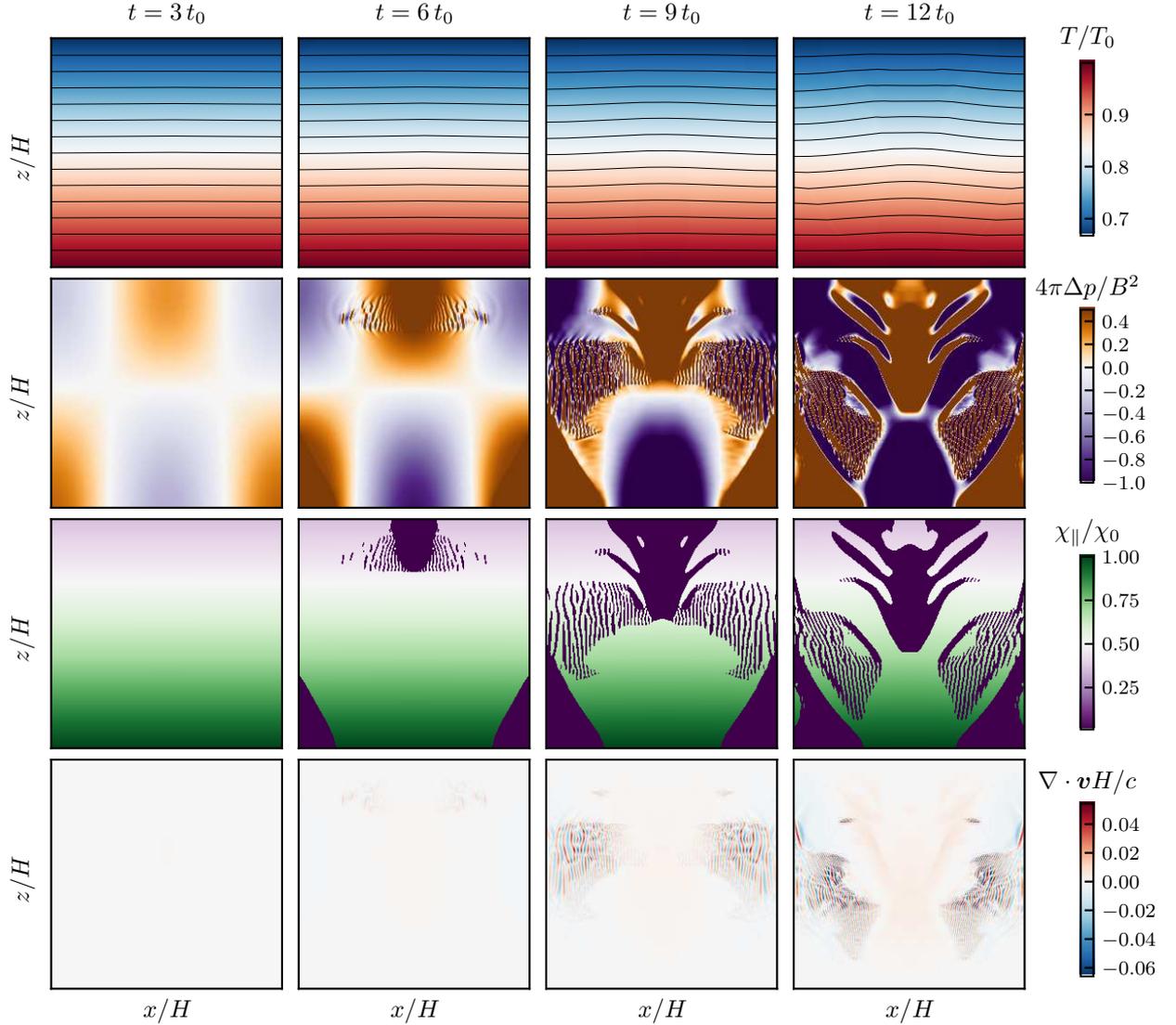}
\caption{
Interruption of the MTI in an \textsc{athena} simulation seeded with an
eigenmode obtained with \textsc{psecas}.  First row:
Temperature evolution
with magnetic field lines (initially horizontal) indicated with solid black
lines.  Second row: Pressure anisotropy attains
large positive values and a
large fraction of the domain becomes mirror unstable and become suppressed by
our subgrid model for heat conduction, see Sect.~\ref{sec:heat-subgrid}.
Third row: The heat conductivity, $\chi_\para$. At later times,
mirror-unstable regions appear as purple patches in which the heat
conductivity is suppressed. Fourth row: Divergence of the velocity field,
which is spatially correlated with striations in the heat conductivity.
The spatial extent shown has size $\HO\times \HO$ where $\HO$ is the scale
height. }
\label{fig:mti_evp_imshow}
\end{figure*}

We illustrate how the linear stage of the MTI can be
interrupted if the heat
conductivity is reduced in mirror-unstable regions. This test of our numerical
setup is performed by
initializing a simulation with an MTI eigenmode.
While the MTI mechanism is
intrinsically local and therefore well described by local, linear
theory (\citealt{Kun11} and Section~\ref{sec:mti-local-theory}),
the $z$-dependence of background variables in our setup (i.e. temperature as well as
density and pressure gradients, see equations~\ref{eq:density-profile}--\ref{eq:pressure-profile}) make the fastest growth rate
obtained using local, linear theory differ by around 20 \% between the top
and bottom of the unstable domain. Consequently, seeding with a local eigenmode does
not give an exact match between theory and simulation.
In order to obtain a completely
clean evolution of an eigenmode, we therefore derive the linearized equations
without making the local approximation in $z$. The resulting eigenvalue
problem is solved for the eigenmodes and corresponding growth rates using
\textsc{psecas} \citep{Berlok2019}. The linearized equations and details on
this procedure, which share features with the quasi-global theory for a
vertical magnetic field \citep{Lat12,Berlok2016b}, are outlined in
Appendix~\ref{app:quasi-global-theory}. For consistency
with the boundary conditions used in the linear theory, the
eigenmode-seeded simulation is performed with reflective boundary conditions and without the stable buffer regions that
we use in the simulations of the nonlinear regime. We note that these boundary conditions cause the eigenmodes to have a larger pressure anisotropy than found for the fastest growing mode
using local linear theory (see details in Appendix~\ref{app:quasi-global-theory}).

We show the time evolution of an exactly seeded simulation where the
background\footnote{This combination of $\beta_0$ and $\Kni$
is chosen such that modification of the MTI evolution occurs already early in the linear stage of the instability. This gives a time interval
of $\approx 5 t_0$ during which
the perturbation amplitude (and corresponding non-linearities)
are small enough that the modifications seen are due to the suppressed
heat conductivity.} has $\beta_0=10^6$ and $\Kni=2000$ in
Fig.~\ref{fig:mti_evp_imshow}.
Here the eigenmode used for seeding the
instability has $k_x H/2\upi = 1$
and growth rate $\sigma t_0 = 0.28305$ where $t_0=\omdy^{-1}=\HO/c_0$
(for comparison, the fastest growing mode has $\sigma t_0 = 0.64700$ and occurs
at $k_x H/2\upi = 13.8374$. See Fig.~\ref{fig:quasi_theoretical_growth_rates} in Appendix~\ref{app:quasi-global-theory}).
The first row of panels in Fig.~\ref{fig:mti_evp_imshow} shows the magnetic field lines (black
solid lines) and temperature evolution (red is hot, blue is cold). The
atmosphere is vertically stratified with temperature that decreases with
height and gravity acts downwards. The system is unstable to the MTI and the
magnetic field lines start to bend due to buoyant motions that grow
exponentially in time.\footnote{This simulation stops early, and the magnetic field
bending is only barely visible at $t=12 t_0$. A nonlinear simulation will be
shown in Section~\ref{sec:nonlinear}.}
The motions
driven by MTI have an associated pressure anisotropy which can excite
microscale instabilities. The  second row of
panels shows $4\upi \Delta p/B^2$ which is a measure of whether the pressure
anisotropy, $\Delta p$, exceeds the mirror threshold ($4\upi \Delta p/B^2 >
1/2$) or the firehose threshold ($4\upi \Delta p/B^2 < -1$).
Mirror-unstable (firehose-unstable) regions appear as saturated orange
(purple) patches in the last three panels.  The heat conductivity, $\chi_\para$, is
reduced by a factor $\Sd=0.01$ in the mirror-unstable regions. Maps of $\chi_\para$, in
 which regions with suppressed thermal
conductivity appear as purple patches, are shown in the third row of panels in
Fig.~\ref{fig:mti_evp_imshow}. Note  here that values for
$\chi_\para$ are
shown normalized by $\chi_0$, the initial value at the bottom of the unstable
domain. The regions with suppressed heat conductivity are predicted by the
quasi-global theory to initially appear at the top center and bottom left- and
right hand sides of the computational domain. This appearance is indeed seen
in the heat conductivity map at $t=6t_0$. The additional high-frequency
striations are not predicted by the linear theory. These features are
spatially correlated with variations in $\del \bcdot \vec{\varv}$, as evident
by comparing the third and fourth rows of Fig.~\ref{fig:mti_evp_imshow}. This
indicates that the small scale variations in heat conductivity are related to
compressible motions of the plasma. These features occur on the grid scale which
make them problematic from a numerical point of view. We therefore dedicate two appendices
(\ref{sec:soundwave} and \ref{app:resolution-study}) in order to ensure that our main
conclusions are unaffected by the striations. We briefly summarize why we believe the
striations occur below.

Sound waves are compressible and can drive a pressure anisotropy by modifying the density.
This means that high amplitude sound waves can trigger the mirror
instability, such that
sound waves, in combination with our subgrid model for heat conduction
(Section~\ref{sec:heat-subgrid}), can lead to a reduction in heat
conductivity.
A study of this
phenomenon, including a theoretical amplitude limit on sound waves
(equation~\ref{eq:int-limit}), is described
in Appendix~\ref{sec:soundwave}.
This study shows that sound wave evolution can be modified and give rise to a
jagged appearance in pressure anisotropy and heat conductivity which looks
remarkably similar to the features seen in the MTI simulation. Given the
visual resemblance and the spatial correlation with the velocity divergence,
we therefore suspect that the striations seen in
Fig.~\ref{fig:mti_evp_imshow} are due to sound waves triggering the
mirror instability.

While excitation of the mirror instability by low-amplitude sound waves in this context is an unwanted consequence of the subgrid model for heat conduction (due to the resulting grid scale features),
\citet{Kunz2020arXiv} have very recently showed that
high-amplitude sound waves can excite the mirror instability in
collisionless systems. Using kinetic analysis and PIC simulations,
they found that the (ion) heat conductivity is suppressed in the
mirror unstable regions.
Further details and discussion about triggering of the mirror instability by sound waves can be
found in Appendix~\ref{sec:soundwave}. In particular, we discuss
how simulations converge despite the grid scale striations in heat conductivity
found in regions with $\del \bcdot \vec{\varv}\neq 0$.

\begin{figure*}
\centering
\includegraphics[trim= 0 12 0 0]{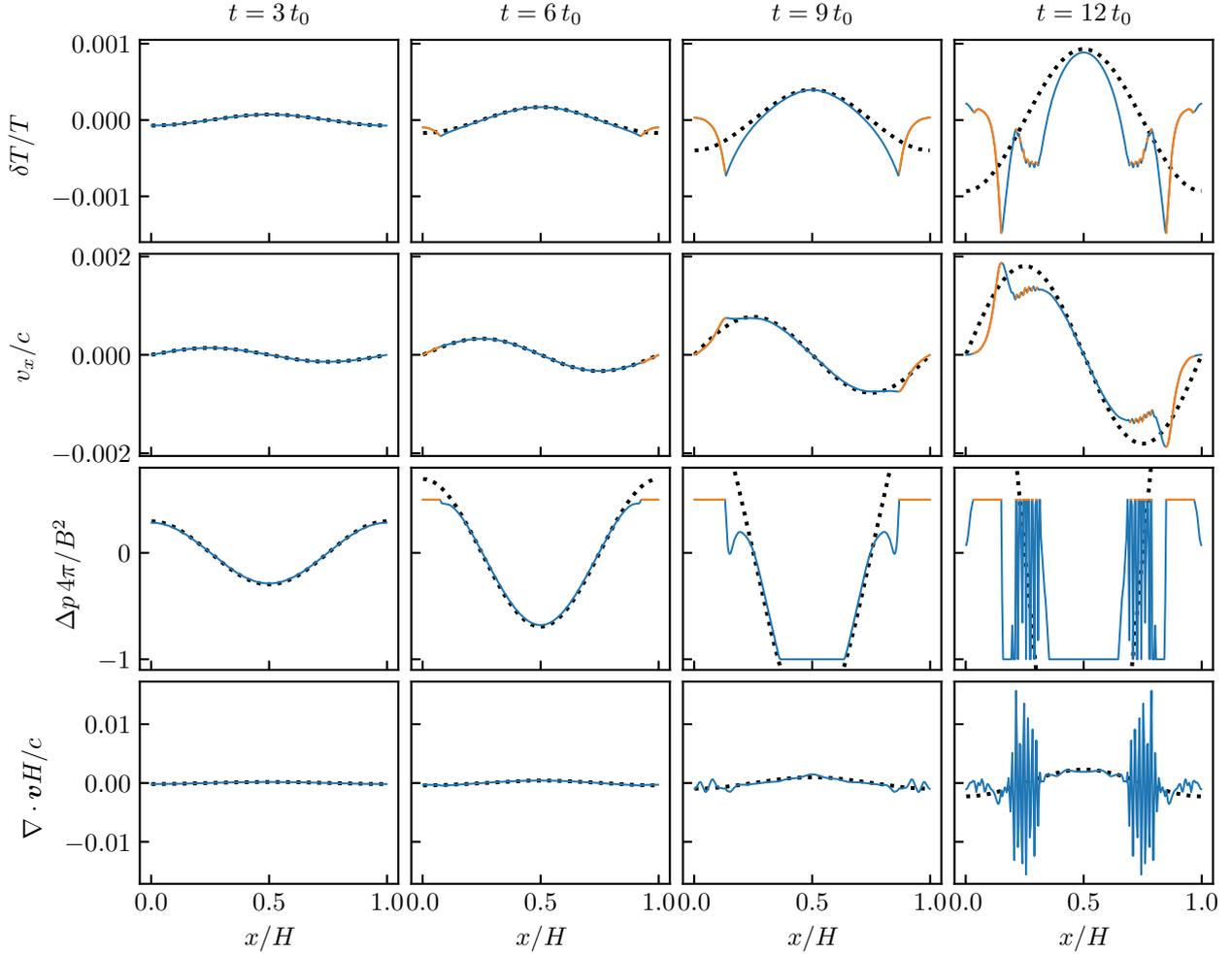}
\caption{Horizontal profiles at $z/L = 1/5$ from
the simulation shown in Fig.~\ref{fig:mti_evp_imshow}. The rows
show $\delta T/T$  (first row), $\varv_x/\cs$
 (second row) , $4\upi \Delta p/B^2$
 (third row), and $\del \bcdot \vec{\varv}H/c$ (fourth row).
The evolution of the \textsc{Athena} simulation (solid lines) is well
described by the \textsc{psecas} eigenmode solution (dotted lines)
until the mirror instability threshold is surpassed. Subsequently,
mirror-unstable regions and resulting
suppression of heat conductivity leads to an interruption of the eigenmode.
}
\label{fig:mti_evp_profile}
\end{figure*}

We now proceed with the analysis of the simulation shown in Fig.~\ref{fig:mti_evp_imshow}
by taking cuts at constant height.
Fig.~\ref{fig:mti_evp_profile} shows cuts at
$z/H=1/5$ and
compares results from the simulation (solid lines) with the linear theory
(dashed lines). The linear theory does not take into account the subgrid
models (nor non-linear terms in the governing equations) and the eigenmode
thus retains its original shape while its amplitude grows exponentially as
time progresses from the left-hand side to the right-hand side in
Fig.~\ref{fig:mti_evp_profile}. We fit the exponential growth in the perturbed
variables and find essentially perfect agreement with the \textsc{psecas}
solution until the onset of suppressed conductivity (see Fig~\ref{fig:mti-linear-fit} in
Appendix~\ref{app:quasi-global-theory}). In the first panel ($t/t_0=3$)
in Fig.~\ref{fig:mti_evp_profile}, the
pressure anisotropy is below the mirror threshold, and the simulation agrees
with the linear theory. At later times, the profiles in the simulation start
to deviate from the sinusoidal shapes of the linear theory. This is because
the pressure anisotropy limiters and associated suppression of heat
conductivity have begun to modify the evolution. The deviation is particularly
evident at $t/t_0=6$ and 9 where the temperature has acquired doubly-peaked
profiles and $\Delta p$ is limited by equation~\eqref{eq:fire_and_mirror}.
Here the mirror-infested regions, in which the heat conductivity is
suppressed, are indicated with orange solid lines. In the last panel
($t/t_0=12$), both the velocity and temperature profile have acquired highly
complex shapes with small regions that spatially fluctuate between being
mirror-infested and retaining full Spitzer conductivity.
Suppression of the thermal conductivity in mirror-unstable
regions thus disrupts the otherwise clean evolution of an MTI eigenmode.
\subsection{Nonlinear simulations}
\label{sec:nonlinear}

We proceed by performing nonlinear simulations which are seeded with Gaussian,
subsonic velocity perturbations with amplitude $10^{-4} \cs$.\footnote{While
we initialize the simulations as a quiescent cluster, we note
that real cluster outskirts are expected to be turbulent due to mergers
and gas accretion. This limitation of our idealized setup is discussed in
Sections~\ref{sec:3D} and \ref{sec:discussion}.}
We compare
simulations with and without heat transport suppression in order to understand
whether the subgrid model for heat transport modifies the saturated state of
the MTI. The simulations have initial magnetic field strength with $\beta_0 =
10^{5}$ and the inverse Knudsen number is $\Kni=2000$.\footnote{The heat
diffusivity and the Braginskii viscosity coefficient are then given by
$\kappa_\para = \chi_\para T/p = 0.012\,T^{5/2}/\rho \ $ and $\nu_\para =
2.4\times10^{-4}\,T^{5/2}/\rho$ in code units, see \citet{Kun12} for details.}
The simulations with and without suppression have $\Sd=0.01$ and $\Sd=1$,
respectively. Note that this is in the regime of $\beta_0$ and $\Kni$
where $\Sd=0.01$ does not significantly change the linear growth rate of the
MTI, see Fig.~\ref{fig:theoretical_growth_rates}.

\begin{figure*}
    \centering
    \includegraphics[trim= 0 15 0 10]{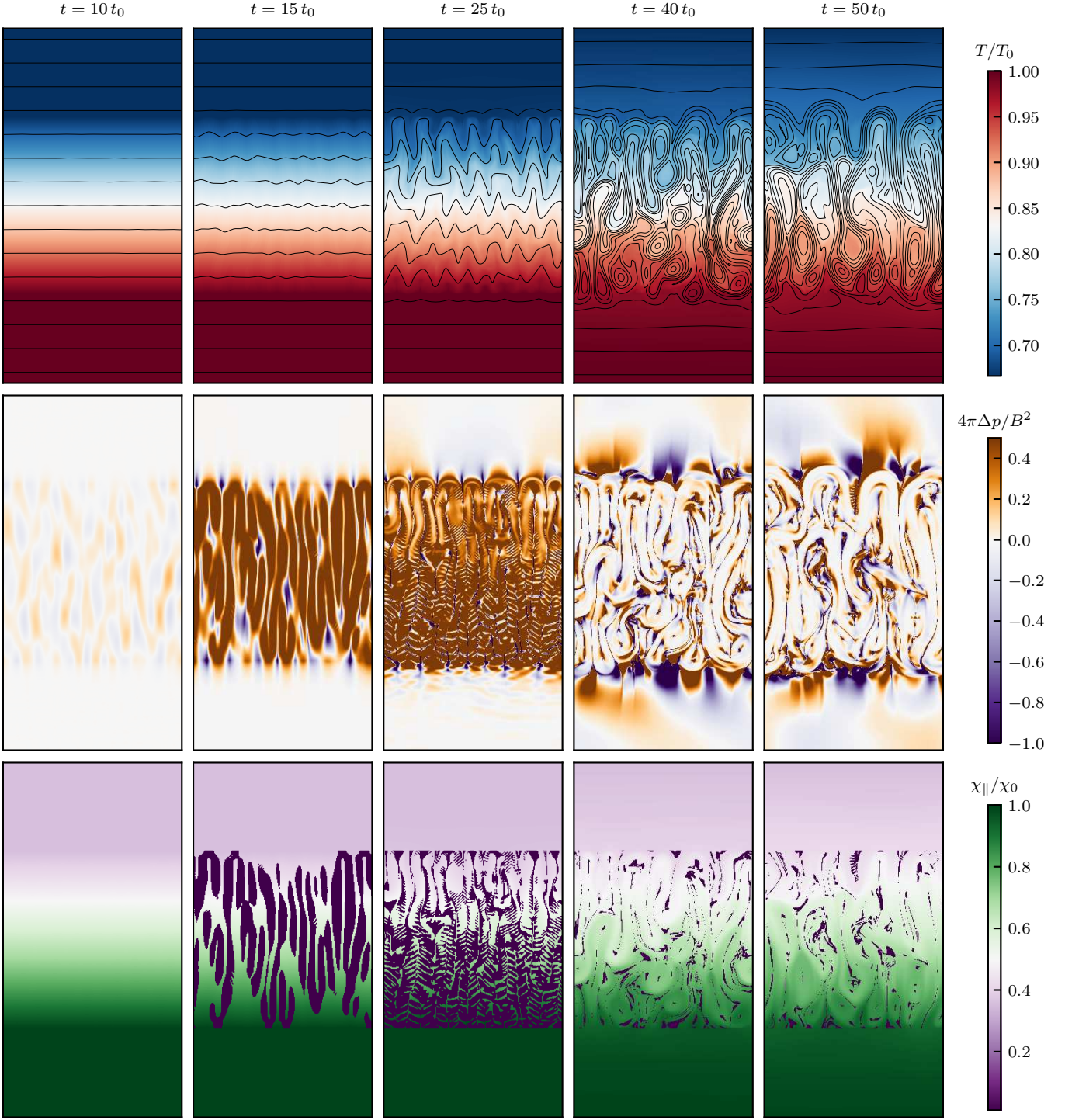}
    \caption{Evolution of temperature and magnetic field (upper row),
    pressure anisotropy (middle row), and heat conductivity (bottom row) in a
    simulation with $\Kni=2000$, $\beta_0=10^5$ and $\Sd=0.01$.
    The computational domain has size $\HO\times 2\HO$ but only the
    middle half of the domain is unstable to the MTI. The top and bottom
    regions are stable, isothermal buffers which shield the central region
    from the reflective boundaries.
    }
    \label{fig:fiducial_simulation}
\end{figure*}

The evolution of the MTI in the simulation with $\Sd=0.01$ is presented in
Fig.~\ref{fig:fiducial_simulation}. The initially horizontal magnetic field
with a temperature gradient pointing downwards leads to the generation of
buoyant motions which change the magnetic field strength and structure
(first row of panels). The inclusion of Braginskii viscosity
suppresses small scale motions and the dominant modes of the MTI grow on
larger scales than in a simulation with only anisotropic heat conduction
\citep{Kun12}. This is the case even though the limiters,
equation~\eqref{eq:fire_and_mirror}, make the simulation less viscous than a
simulation without them (the difference in magnetic field line
morphology between simulations with unlimited Braginskii viscosity,
limited Braginskii viscosity, and zero viscosity is clearly illustrated in
figure 17 in \citealt{Kun12}, see also figure 5 in \citealt{Berlok2016b}).

The plasma motions created by the MTI
in turn lead to the generation of a non-zero pressure anisotropy, as given by
equation~\eqref{eq:pressure_anisotropy}. We show the spatial evolution of the
pressure anisotropy in the middle row
and the corresponding heat conductivity in the bottom row of
Fig.~\ref{fig:fiducial_simulation}.
 At $t/t_0 = 10$, the pressure
anisotropy  is still
below the mirror instability threshold and the heat conductivity takes the full
Spitzer value. At $t/t_0 = 15$, the mirror threshold has been exceeded
and purple large-scale regions of suppressed heat conductivity appear.
At $t/t_0 = 25$, the purple patches have progressed to smaller scales with
striations similar to the ones found in Fig.~\ref{fig:mti_evp_imshow}.
We again find that these features are spatially correlated with the velocity
divergence (not shown).

At $t/t_0 = 25$, roughly fifty percent of the computational domain is unstable
to the mirror instability according to the linear stability criterion. Near
the end of the simulation (fourth and fifth panels of
Fig.~\ref{fig:fiducial_simulation}, at $t/t_0=40$
and 50), however, less than 10 per cent
of the computational domain is unstable according to the linear stability
criterion for the mirror instability. The low fraction of the mirror unstable
volume is probably related to the growth in magnetic field strength. While
magnetic field growth is necessary in order to generate positive
pressure anisotropies (which will lead to triggering of the mirror
instability) an increased magnetic field strength also decreases the
plasma-$\beta$. This can have the side-effect that the increased magnetic
field strength ends up stabilizing the mirror instability. That is, $B^2/8\upi
< \Delta p \propto \mathrm{d}B/\mathrm{d}t $ becomes more difficult to achieve
as $B$ grows in magnitude and $\mathrm{d}B/\mathrm{d}t$ saturates.

\begin{figure*}
    \centering
    \includegraphics[trim= 0 20 0 0]{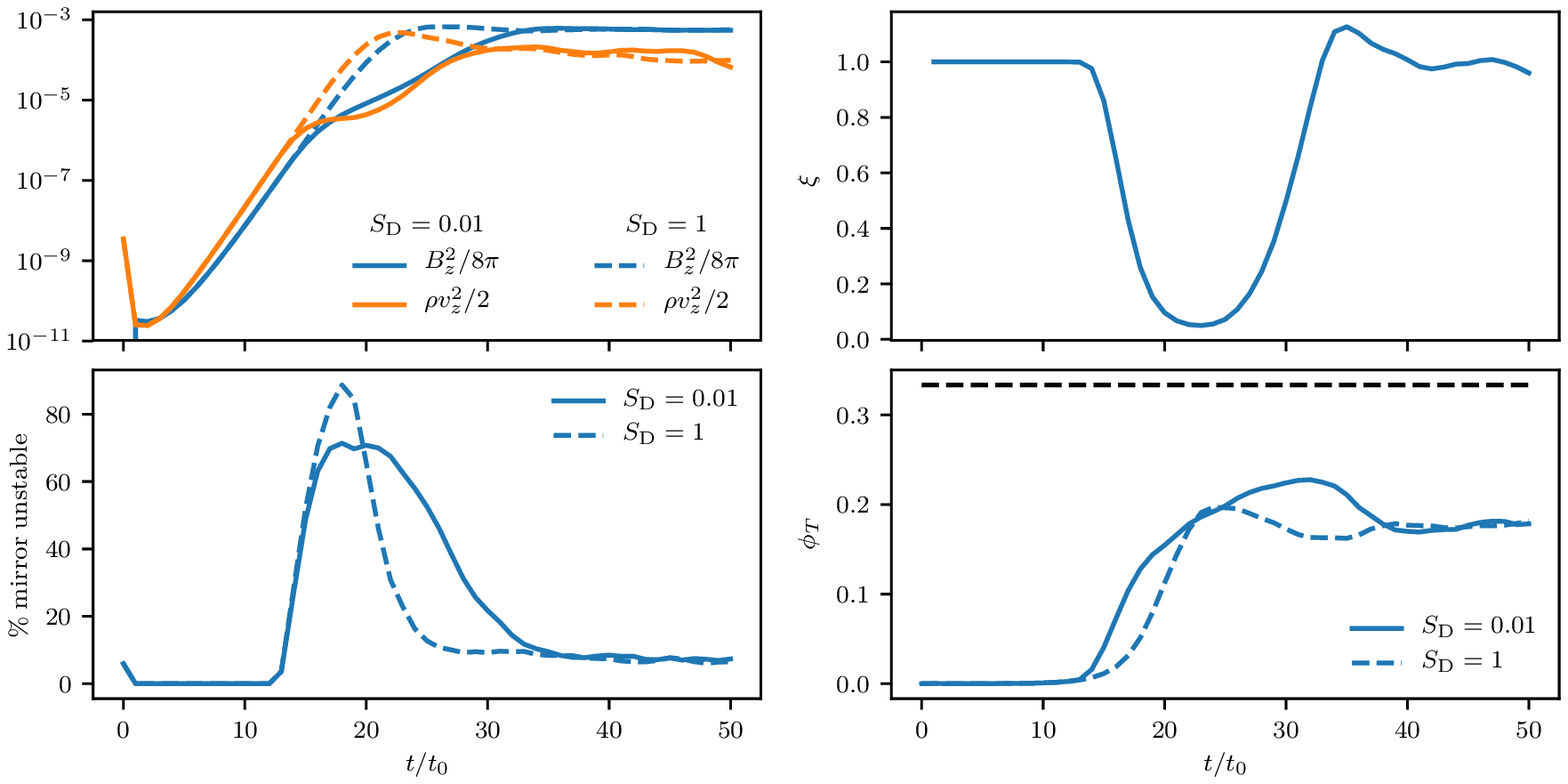}
    \caption{Time evolution in simulations of the MTI with $\Sd=0.01$ (solid)
    and $\Sd=1$ (dashed). Upper left: Vertical magnetic and kinetic
    energies. Upper right: Ratio of magnetic energy in the $\Sd=0.01$
    and $\Sd=1$ simulations, $\xi$, defined in equation~\eqref{eq:xi_def}. Lower left:
    Mirror unstable volume fraction. Lower right: Normalized, magnetic
    field-aligned temperature gradient, $\phi_T$, defined in
    equation~\eqref{eq:phi_T}.}
\label{fig:fiducial_lineplot}
\end{figure*}

In a simulation seeded
with Gaussian velocity components,
the evolution of the MTI is
such that the plasma almost entirely avoids becoming unstable to the firehose
instability by arranging magnetic field lines to lie across velocity
gradients, see \cite{Kun12}. We have checked this by using the instability
criterion for the firehose, see equation~\eqref{eq:fire_and_mirror}
and the maps of $4\upi\Delta p/B^2$ in the middle row of
Fig.~\ref{fig:fiducial_simulation}. Note that this differs from a
simulation seeded with an MTI eigenmode where
a significant region of space becomes firehose unstable
(see Fig.~\ref{fig:mti_evp_imshow}).
We analyze the simulations quantitatively by considering the time evolution of
a number of key quantities.

The upper, left-hand panel of Fig.~\ref{fig:fiducial_lineplot} shows the
vertical kinetic and magnetic energies, $\langle \rho \varv_z^2/2 \rangle$ and
$\langle B_z^2/8\upi \rangle$, where $\langle \rangle$ denotes an average over
the unstable part of the computational domain ($0.5<z/H<1.5$). The simulation
with $\Sd=0.01$ starts to deviate from the reference simulation with $\Sd=1$ at
around $t/t_0=15$. This deviation occurs because a substantial fraction of the
volume is mirror-unstable, so that the effective heat conductivity is
suppressed, inhibiting the growth. The deviation between the simulation with
$\Sd=0.01$ and the one with $\Sd=1$ is however only temporary. In the end, the
energies in the saturated states are roughly similar.

In order to further quantify this surprising result, we define $\xi$ to be the
ratio of the vertical magnetic energy in $\Sd=0.01$ and $\Sd=1$ simulations, i.e.,
\be
    \label{eq:xi_def}
    \xi \equiv \f{\langle B_z^2 (\Sd=0.01)\rangle}
    {\langle B_z^2 (\Sd=1)\rangle} \ .
\en
The advantage gained by considering $\xi$ instead of the energies is that
$\xi$ does not vary by many orders of magnitude. We show the evolution of
$\xi$ in the upper right-hand panel of Fig.~\ref{fig:fiducial_lineplot}. While
the value dips down to less than 10 per cent at around $t/t_0=25$ it saturates with
a value close to unity in the end.

We also consider the fraction of the computational domain that is unstable to
the mirror instability (lower, left-hand panel of
Fig.~\ref{fig:fiducial_lineplot}). Due to the subgrid model employed, this is
a measure of the reduction of the heat conductivity. We observe that the start
and end of the dip in $\xi$ are roughly coincident in time with the peak in
and subsequent decay of the mirror unstable fraction.

Finally we consider the volume average of the absolute value of the
field-aligned temperature gradient, i.e.
\be
    \label{eq:phi_T}
    \phi_T = \left\langle \left|\b \cdot \nabla T\right|\f{H}{T}\right\rangle .
\en
At first sight, $\phi_T$ might appear to be a purely geometrical construct but
as we will argue here, $\phi_T$ is a measure of the efficiency of heat
transport along magnetic field lines. The minimum value of $\phi_T$ is zero
and occurs in the limit of infinitely fast heat conduction where closed
magnetic field lines become isotherms. In the opposite limit, where heat
conduction is absent, the expected upper bound on the value of $\phi_T$ is the
one found with a magnetic field everywhere perfectly aligned with the initial
temperature gradient ($\b=\ez$). Using
equation~\eqref{eq:temperature-profile}, we find this value to be
$\phi_T=1/3$. Hence, we expect $\phi_T$ to be large in simulations where the
heat conductivity is low (either due to a large $\Kni$ or to a high percentage
of the domain being unstable to the mirror instability) and small in
simulations where the heat conductivity is fast.

We show the time evolution of $\phi_T$ in the lower, right-hand panel of
Fig.~\ref{fig:fiducial_lineplot}, again comparing with the reference
simulation. Due to the suppression of $\chi_\para$, the simulation with
$\Sd=0.01$ attains a somewhat higher value of $\phi_T$ than the reference
simulation with $\Sd=1$. However, this only occurs during the period of time
where a significant volume fraction is mirror-unstable (compare the lower
left-hand panel of Fig.~\ref{fig:fiducial_lineplot} with the lower right-hand
panel).

\subsection{Parameter study}
\label{sec:parameter_study}

\begin{figure*}
\includegraphics[trim = 0 20 0 0]{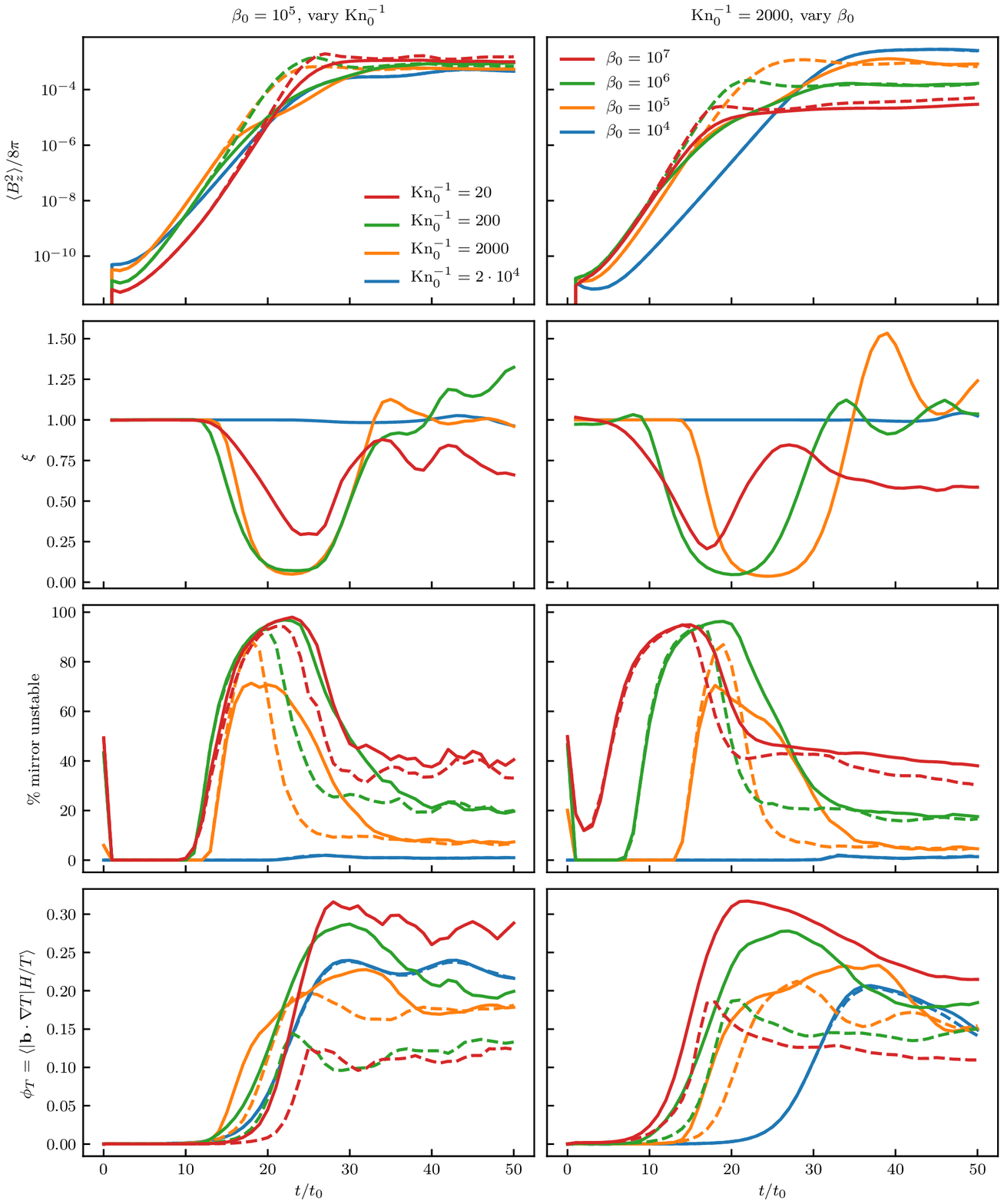}
\caption{The effect of mirror-suppressed heat conductivity
at various values of $\beta_0$ and $\Kni$. We show the
evolution of vertical magnetic energy (first row), ratio of magnetic energy in
simulations with and without suppression, $\xi$, (second row, see equation
\ref{eq:xi_def}), mirror-unstable fraction (third row) and field aligned
temperature gradient, $\phi_T$ (fourth row, see equation \ref{eq:phi_T}).
Solid lines have suppression factor
$\Sd=0.01$ and dashed lines are reference simulations with $\Sd=1$.
As expected from Equation~\ref{eq:mirror-Knbeta}, the mirror-unstable
fraction is largest in the simulations with the largest $\beta_0\Kn$
($\beta_0\Kn=5\times10^3$, the red lines). These simulations thus have the
largest regions with suppressed
thermal conductivity and $\phi_T$ becomes larger than in the simulations without
suppression. Despite this change in conductivity, the final saturation energy
of the MTI is ultimately only slightly modified. We interpret this latter
observation using linear theory which, in the limit $\beta_0 \Kn \gg1$, shows
that the MTI growth rates remain significant even when the thermal conductivity
is suppressed (see Fig.~\ref{fig:theoretical_growth_rates}).  }
\label{fig:parameter_plot}
\end{figure*}

In the previous section we found that the MTI is quite robust to suppression
of heat conductivity, at least when $\beta_0=10^5$ and $\Kni =2000$. Here we
perform a parameter study in order to see whether there are other parameter
regimes where the MTI is more severely modified.
We vary the collisionality, via $\Kni$ at fixed $\beta_0=10^5$ (left-hand
column of Fig.~\ref{fig:parameter_plot}) and the magnetic field strength, via
$\beta_0$, at fixed $\Kni=2000$ (right-hand column of
Fig.~\ref{fig:parameter_plot}). We show in Fig.~\ref{fig:parameter_plot} the
evolution of vertical magnetic energy (first row), the ratio of magnetic
energy in simulations with and without suppression, $\xi$, (second row, see
equation \ref{eq:xi_def}), the mirror-unstable fraction (third row) and the
field-aligned temperature gradient, $\phi_T$ (fourth row, see equation
\ref{eq:phi_T}). The solid lines correspond to simulations with heat transport
suppression ($\Sd=0.01$) and the dashed lines are references with full Spitzer
conductivity ($\Sd=1$).

The left- and right-hand panels of Fig~\ref{fig:parameter_plot} have many
features in common. This is because the important parameter for mirror
instability (see equation \ref{eq:mirror-Knbeta}), $\beta_0\Kn$, only takes
four different values, i.e., $\beta_0\Kn = 5$, $50$, $500$ and $5000$.
For $\beta_0\Kn = 5$ (shown with blue lines in Fig.~\ref{fig:parameter_plot}),
the simulation with suppression is virtually indistinguishable from the
reference simulation. The explanation is simply that the mirror-instability
threshold is almost never surpassed in these simulations (third row in
Fig.~\ref{fig:parameter_plot}).
For $\beta_0\Kn = 50$ (shown with orange lines in
Fig.~\ref{fig:parameter_plot}), a very significant volume fraction ($>60$\%)
is mirror-unstable at $t/t_0\approx 20$ but the fraction goes down to 10\% at
the end of the simulation. This is evidently not enough to effectively
suppress heat transport (i.e., the deviation in $\phi_T$ is temporary) and the
saturated energies in the $\Sd=0.01$ simulation are similar to the reference
simulation. When $\beta_0\Kn=500$ (green lines), the final mirror-unstable
fraction is $\approx 20$\% and the saturated energies still do not differ by
much between the $\Sd=0.01$ and $\Sd=1$ simulations. The $\xi$ parameter in fact
indicates that there is $\approx25$\% more energy in the $\Sd=0.01$ simulation
with $\Kni=200$. This is however within the uncertainty given by a finite
numerical resolution (see Appendix~\ref{app:resolution-study} for a
convergence study and Tables~\ref{tab:simulations} and
\ref{tab:convergence-simulations} for an overview of the numerical resolutions
used in our simulations). Finally, for an extreme value of $\beta_0\Kn = 5000$, the
mirror-unstable fraction remains above 40 \% at the end of the simulation, and
the simulations with $\Sd=0.01$ saturate with slightly less energy than the
reference simulation ($\xi\approx3/4$ at the end).

The trends outlined above can be understood using the MTI linear theory (see
Section~\ref{sec:mti-local-theory}). Here we found that systems with large
$\beta_0$ and small $\Kni$ (i.e. large $\beta_0 \Kn$) to be the most
susceptible to the mirror instability and to have the largest MTI growth rates
(when calculated using the full Spitzer conductivity). In this regime, the
fastest mode of the MTI is so deeply in the fast heat conduction limit that
suppression of the heat conductivity only slightly lowers the growth rate (see
Fig.~\ref{fig:theoretical_growth_rates}). As a consequence, the
difference in saturated energies between simulations with $\Sd=0.01$ and
references with $\Sd=1$ differ by less than a factor of $\sim 2$ over all
of the investigated parameter space.

\begin{figure*}
\includegraphics[trim= 0 20 0 0]{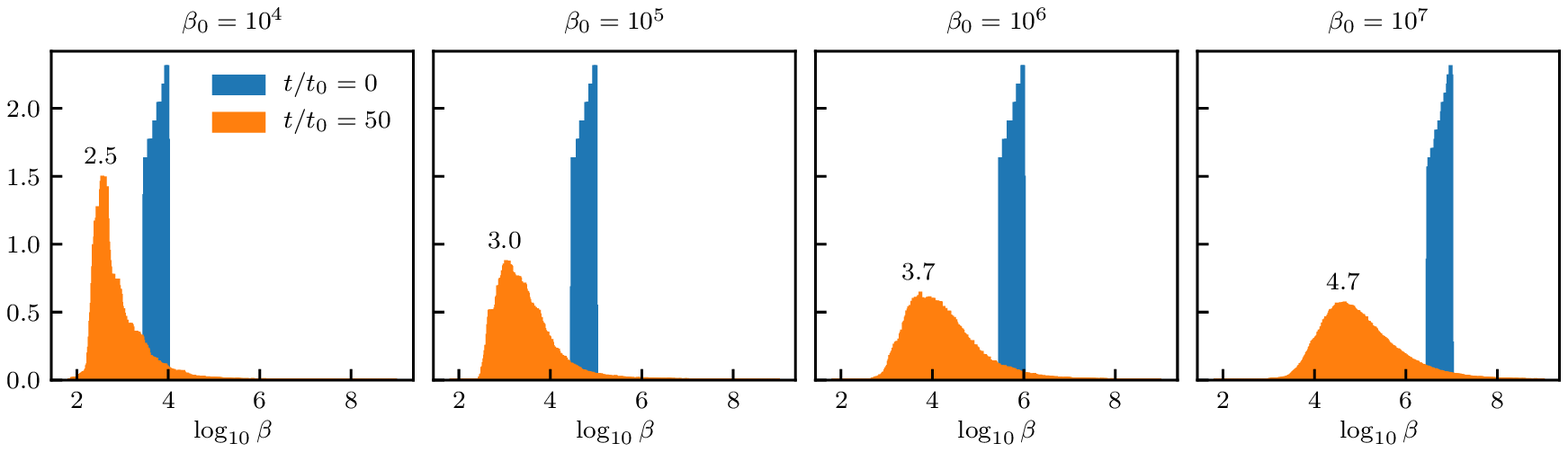}
\caption{Distribution of $\beta$ in the central, unstable part of the domain
at $t/t_0=0$ (blue) and $t/t_0=50$ (orange) in simulations with
different initial magnetic field strengths. The MTI increases the magnetic
field strength and thus decreases $\beta$. The modes of the distributions
are indicated at $t/t_0=50$.}
\label{fig:beta-histograms}
\end{figure*}

For the ICM,  the estimated values are $\beta\sim 10^2$--$10^4$
and $\mathrm{Kn}^{-1}\sim10$--$100$
in the outskirts of galaxy clusters \citep{Car02,Vik06}.
The physically relevant regime for $\beta \mathrm{Kn}$
is therefore $\sim 1$--$10^3$. Since simulations with small values of $\Kni$
are computationally expensive (due to the time step constraint associated with
thermal transport), we chose for practical reasons $\Kni=20-2\times10^4$. This
allowed us to perform very high resolution simulations for $\Kni=2000$, a
necessity for ensuring that our findings have numerically converged
(see resolution study and considerations in
Appendix~\ref{app:resolution-study}). In order
to capture the physically relevant regime for $\beta \mathrm{Kn} \sim
1$--$10^3$, we increased the values for $\beta_0$ as well, i.e.,
$\beta_0=10^4$--$10^7$ in the right-hand column of
Fig.~\ref{fig:parameter_plot}. Our initial conditions
then cover the range $\beta_0 \Kn = 5$--5000. During the evolution of the simulation,
the magnetic field strength is amplified by the MTI. This decreases the mean value of $\beta$
such that the simulations with $\beta_0=10^4$--$10^7$ at $t/t_0=0$ have $\beta=10^{2.5}$--$10^{4.7}$ as the most frequent values at $t/t_0=50$ (see Fig.~\ref{fig:beta-histograms}). The simulations
with $\beta_0=10^5$ thus have $\beta$ in the relevant range, $\beta= 10^2$--$10^4$, in more than 80 per cent of the MTI unstable domain at $t/t_0=50$ (with 95 per cent for $\beta_0=10^4$).

For the parameter scan, the important parameter for mirror instability, $\beta_0 \mathrm{Kn}_0$,
was chosen to be in the physically relevant range for the outskirts of galaxy clusters. And while we find modifications to the linear regime of the instability
(i.e. decreased growth rates),
differences are not present or are only moderate in the nonlinear regime.
Thus over a factor of $\sim 10^3$ in the product of thermal-to-magnetic pressure
ratio, $\beta_0$, and collisionality measure, $\mathrm{Kn}_0$,
only a modest change to the saturation of the MTI is found by
taking into account suppression of heat conductivity by the mirror instability.
\subsection{3D simulations}
\label{sec:3D}

\begin{figure}
\centering
\includegraphics[width=0.9\columnwidth]{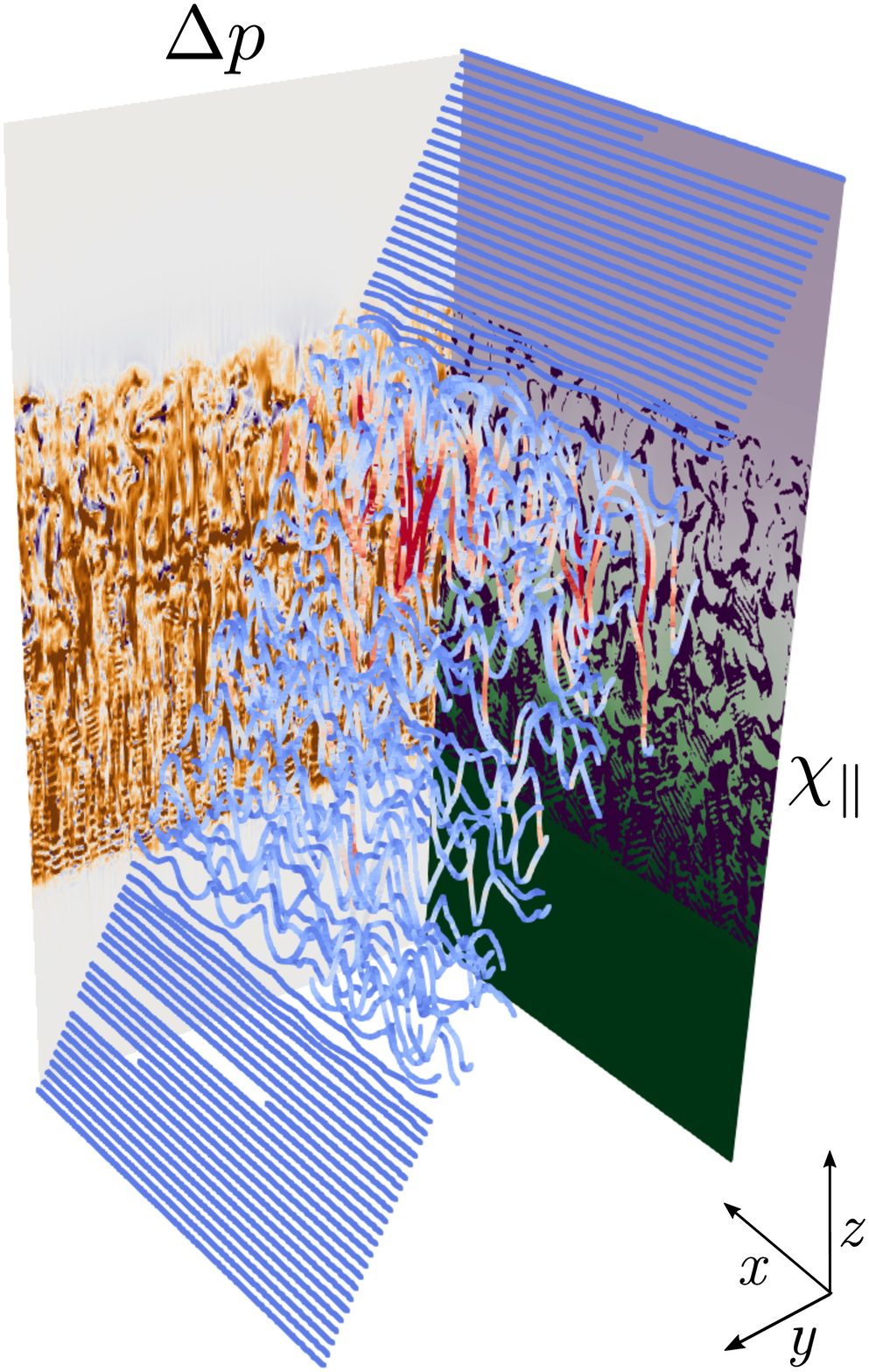}
\caption{3D MTI simulation with $\Sd=0.01$ at $t/t_0=28$.
The magnetic field lines are colored with red (blue) indicating higher
(lower) magnetic field strength.
field. The vertical slices show the pressure anisotropy, $\Delta p$,
and the heat conductivity, $\chi_\para$. Mirror-unstable regions
appear as purple patches in $\chi_\para$. See
Fig.~\ref{fig:fiducial_simulation} for the 2D version.}
\label{fig:3D-MTI-vis}
\end{figure}

We have so far only presented 2D simulations of the MTI, which are less
computationally intensive than their more realistic 3D counterparts.  The
smaller computational cost allowed us to perform a parameter study at high
numerical resolution in the previous section. In order to ensure that our
findings are not an artifact of the reduced dimensionality, we also perform
four 3D simulations.  We consider the case $\beta_0=10^5$ and $\Kni=2000$
only. Initially, the magnetic field is oriented along the $x$-direction and
gravity acts in the negative $z$-direction. The extra dimensionality of the
simulation now allows motions in the horizontal $y$-direction and all three
velocity components are initialized with Gaussian noise.

We present a visualization of a simulation with $\Sd=0.01$ at $t/t_0=28$ in
Fig.~\ref{fig:3D-MTI-vis}. Note that this figure shows the full computational
domain, including the buffer regions used below and above the unstable region.
The slices show pressure anisotropy and heat conductivity, respectively
(compare with Fig.~\ref{fig:fiducial_simulation}).  The magnetic field lines
display signatures of buoyant motions inside the central region.

\begin{figure}
\includegraphics[trim= 0 25 0 0]{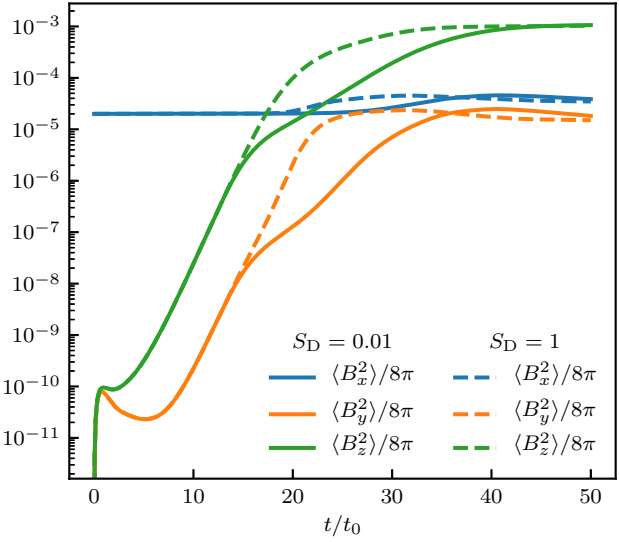}
\caption{Evolution of magnetic energies in the 3D MTI simulation shown
in Fig.~\ref{fig:3D-MTI-vis}. Numerical resolution is $256\times256\times512$
and the same qualitative picture is found at resolution
$128\times128\times256$.}
\label{fig:energy_3D_highres}
\end{figure}

We show the evolution of the volume average of the three magnetic energy
components in a 3D simulation with $\Sd=0.01$ (solid lines) and $\Sd=1$
(dashed lines) in Fig.~\ref{fig:energy_3D_highres}. As for the 2D simulation
presented in Fig.~\ref{fig:fiducial_lineplot}, suppression of heat transport
in mirror-unstable regions causes the growth rate to decrease. Nevertheless,
the saturated energy levels are not changed by employing the subgrid model for
heat transport described in Section~\ref{sec:heat-subgrid}.

As evident from Figs.~\ref{fig:3D-MTI-vis} and \ref{fig:energy_3D_highres},
the motions are primarily in the $z$-direction and while a significant $B_z$
component is created, the $B_y$ component remains sub-dominant. This is
likely the reason why we find the evolution of vertical magnetic energy in
the $\Sd=0.01$ and $\Sd=1$ simulations to be almost identical to that in the
2D simulations (compare Figs.~\ref{fig:fiducial_lineplot} and
\ref{fig:energy_3D_highres}). That is, the saturation level of magnetic
energy is the same in simulations with and without heat transport
suppression, regardless of whether a 2D or a 3D study is performed.
In addition, we find the same behavior for other
diagnostics when comparing 2D and 3D simulations (such as the time-evolution
of the mirror unstable fraction).

\begin{figure}
\includegraphics[trim= 0 25 0 0]{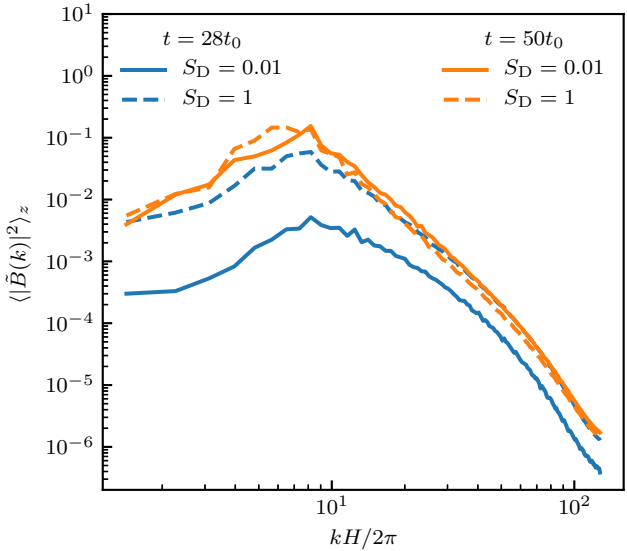}
\caption{Magnetic field power spectrum averaged over the unstable domain.
At $t=28t_0$,
the energy in the simulation with $S_\mathrm{D}=0.01$ (solid blue line) lags
behind the reference simulation with $S_\mathrm{D}=1$ (dashed blue line), in particular on
large scales across which heat conduction is suppressed at this point in
time. At $t=50t_0$, the power in the simulation including suppression
by the mirror instability has caught up with the reference simulation on
all scales (solid and dashed orange lines coincide).}
\label{fig:power_3D_highres}
\end{figure}

Finally, we study the time evolution of magnetic energy as a
function of scale in the 3D simulations. Since the simulation
domain is not periodic in $z$, we calculate $|B(k)|^2$ in each
horizontal slice of the unstable domain (here $k = \sqrt{k_x^2 + k_y^2}$ is
the horizontal wavenumber) and perform a vertical average in $z$.
The resulting profiles are shown in Fig.~\ref{fig:power_3D_highres} with
blue (orange) lines at $t=28t_0$ and solid (dashed) lines obtained from
simulations with $\Sd=0.01$ ($\Sd=1$). At $t=28t_0$ the simulation
with suppression of heat conduction is lagging behind the reference
simulation at all scales. The difference is most pronounced across large
scales where heat conduction is more efficiently suppressed. At $t=50t_0$
this difference between the $\Sd=0.01$ and $\Sd=1$ simulations has however
completely disappeared, and the simulations have the same energy both on
large and small scales. The finding that the saturated state is insensitive
to heat conduction suppression thus holds both in terms of volume-averaged
energies (Fig.~\ref{fig:energy_3D_highres}) and on a scale-by-scale basis
(Fig.~\ref{fig:power_3D_highres}).
The latter finding has importance for understanding how the MTI operates
in combination with external turbulence. In particular,
\citealt{2011MNRAS.413.1295M} found that the MTI can maintain its power
spectrum on scales larger than the external driving scale even when
strong external turbulence swamps outs MTI turbulence on smaller
scales (see their
figure 14). Large scale MTI-driven motions thus survive both in simulations
with external turbulence \citep{2011MNRAS.413.1295M} and in our simulations
with $\Sd=0.01$. Understanding the combined influence of these two
effects is outside the scope of the present work but we briefly discuss
what could happen in Section~\ref{sec:discussion}.

In contrast to \citet{2011MNRAS.413.1295M},
\citet{Ruszkowski2011ApJ} studied the interaction between external turbulence
and the MTI by performing cosmological simulations
of cluster formation including anisotropic transport.
They found little difference in radial bias of magnetic fields
between simulations with
and without anisotropic heat conduction. This could indicate that external
turbulence suppresses the MTI. However, \citet{Ruszkowski2011ApJ}
caution that \emph{i)} it is difficult to disentangle the effect of the MTI and
radial flow on the orientation of the magnetic fields \emph{ii)}
a systematic study of several galaxy clusters is required
to make a definite statement regarding relative contributions of
MTI/external turbulence
as the ICM turbulence properties will differ from cluster to cluster.
In addition, \citet{2011MNRAS.413.1295M}
show in their idealized setup that both external turbulence and the MTI lead
to a randomization of the magnetic field orientation (see their figure 13).
This makes volume integrated
quantities poor diagnostics for determining the importance of the
MTI. As such, a detailed understanding of
whether the MTI is playing an active role in the simulations of
\citet{Ruszkowski2011ApJ}
might require spectral analysis such as the one performed
by \citet{2011MNRAS.413.1295M}.

Numerical studies of turbulence in weakly collisional plasmas
have so far been limited to idealized, turbulent box setups (e.g.
\citealt{Squire2019JPlPh,Kempski2019MNRAS}).
Most notably \citet{Squire2019JPlPh}
have found that turbulence with Braginskii viscosity can display
so-called \emph{magneto-immutability}, i.e. that the turbulent motions are
constrained to those that do not generate pressure anisotropy. While this
pertains in particular to very viscous simulations without pressure anisotropy
limiters (equation~\ref{eq:fire_and_mirror}), even less viscous simulations with the limiters
applied display a narrower distribution in the parallel rate-of-strain
($bb:\nabla v$, which in their incompressible simulations is directly
proportional to $\Delta p$). Of primary interest to the present work, the
anisotropic viscous flux acts to reduce $\Delta p$ and thus also the
fraction of the volume which is unstable to the firehose and mirror
instabilities (see figure 4 in \citealt{Squire2019JPlPh}).
Similar conclusions have been found in a
study of the weakly collisional magnetic dynamo (\citealt{St-Onge2020},
see in particular their
appendix B1 and figure 23) and in
a study of turbulence driven by the magneto-rotational instability
in weakly collisional accretion disk (\citealt{Kempski2019MNRAS}, see their
figure 9).
The studies by \citet{Squire2019JPlPh,St-Onge2020}
considered incompressible turbulence
using an Ornstein-Uhlenbeck process for the forcing term. Quantifying the
mirror (and firehose) unstable fraction(s) in turbulence driven in a
cosmological setting, i.e., where the turbulence is (at least weakly)
compressible and driven by
large-scale accretion and mergers has yet to be done and is an interesting
avenue for future research.

\section{Summary and Discussion}
\label{sec:discussion}

The MTI has been theorized to drive turbulence in the outer regions of galaxy
clusters. In order to operate, the MTI requires fast heat conduction directed
along the direction of the magnetic field \citep{Bal00,Bal01}. This type of
anisotropic heat transport is expected in the ICM according to \citet{Bra}
because the high temperature and low density of the ICM makes Coulomb
collisions rare. Particles are therefore expected to be effectively tied to
the magnetic field, leading to transport of heat and momentum primarily along
the field. This picture is muddled by various microscale instabilities which
are not taken into account  in the classical transport theory. PIC simulations
show that microscale instabilities can create electromagnetic fluctuations
that can scatter particles, effectively increasing the collision frequency and
changing the heat transport properties of the plasma. In particular,
\citet{Kom16} found that the ion mirror instability can suppress the heat
conductivity by a factor $\Sd=0.2$. We use their results to motivate a subgrid
model for the heat conductivity in Braginskii-MHD simulations. Our subgrid
model is simple: In regions where the mirror instability threshold is
exceeded, $\Delta p/p > \beta^{-1}$, we use a reduced heat conductivity,
$\chi_\mathrm{eff}=\Sd\chi_\para$. Whereas \citet{Kom16}
found $\Sd=0.2$, we have exaggerated mirror-suppression of heat transport by
setting $\Sd=0.01$ throughout the main body of the paper.

Since Braginskii-MHD simulations are less
computationally intensive than PIC simulations, this  subgrid method allows us to
perform simulations that model a larger spatial extent of the ICM compared to
what would be possible with PIC simulations.
Using this approach, we perform a series of simulations of the MTI
in both 2D (Sections~\ref{sec:mti-interrup}, \ref{sec:nonlinear} and
\ref{sec:parameter_study}) and 3D (Section~\ref{sec:3D}).
    In these simulations, the MTI grows exponentially until its
amplitude is large enough to exceed the threshold for stability of the mirror
instability. When this occurs, the heat conductivity is suddenly dramatically
reduced in a large volume of the system. One would naively expect this to
quench the MTI which  depends on efficient heat
transport along magnetic field lines. However, this intuition turns out to be
wrong and we have found surprisingly small  effects on the evolution of the MTI.

We explain this behavior as follows:
Whether the mirror instability is active in large regions of the
simulation depends on the value of $\beta_0 \Kn$ where $\beta_0$ is the ratio
of thermal to magnetic pressure and $\Kn$ is the Knudsen number (see
Section~\ref{sec:coeffs}).
When $\beta_0 \Kn=5$, the MTI-generated motions are not sufficient to excite the
 mirror-instability , while it is excited when $\beta_0 \Kn=5000$
(see Fig.~\ref{fig:parameter_plot}).  The associated
heat transport suppression does
not, however, quench the MTI in the latter case because heat conduction
remains fast even after
suppression when $\beta_0 \Kn=5000$. We show this by calculating the
MTI growth rates using linear theory in Section~\ref{sec:mti-local-theory},
see Fig.~\ref{fig:theoretical_growth_rates}.
\begin{figure}
    \centering
    \includegraphics[trim= 0 15 0 0]{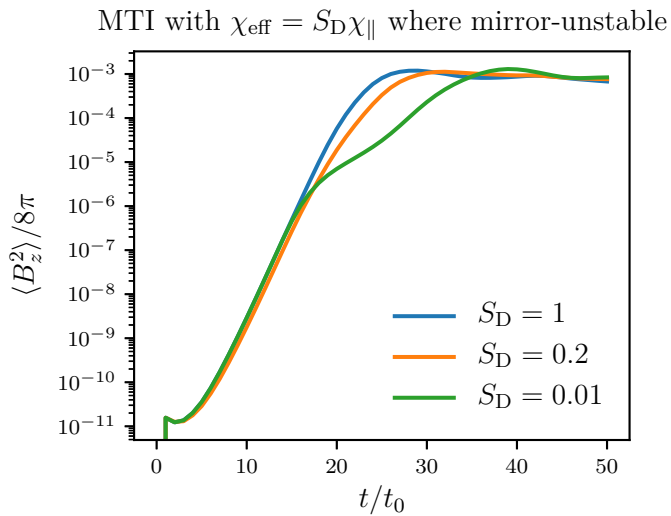}
    \caption{Evolution of energies in simulations with varying
    suppression factor, $S_\mathrm{D}$, at fixed $\Kni = 2000$ and
    $\beta_0 = 10^{5}$. The growth rate of the MTI is reduced in
    the simulations with $\Sd<1$ but the saturated energies are not.}
\label{fig:energies_vary_factor}
\end{figure}

In summary, when $\beta_0 \Kn$ is small, the mirror-instability does not
modify the MTI-evolution (simply because it is not active) and when $\beta_0
\Kn$ is large, the MTI is so far into the fast heat conduction regime, that
suppression of the thermal conductivity, even by a factor $\Sd=0.01$,
is insufficient to severely limit
its growth rate. For a range of values of $\Kni$ and
$\beta$, the evolution of
the MTI therefore appears to be quite robust to reductions of the heat
conductivity by the mirror instability.
And while we do not find a large effect on the MTI evolution when $\Sd=0.01$ and
$\beta_0 \Kn$ is large, the effect becomes even smaller when the value
inferred by \citet{Kom16}, $\Sd=0.2$, is used. This is illustrated in
 Fig.~\ref{fig:energies_vary_factor} where we compare
 simulations with $\Sd=0.01$,
0.2 and 1 and find almost identical evolution for $\Sd=0.2$ and 1.

There are, however, limitations to the present study.
 First, we have only
considered plasma motions generated by the MTI itself. This
yields an interesting dynamical system with negative feedback, i.e., fast heat
transport drives vigorous plasma motions, which suppress the heat transport,
which then only drives weaker motions and allows the heat transport to again
increase (thus closing the loop). It is however important to mention that several
other mechanisms generate plasma motions and associated pressure-anisotropy,
e.g., AGN and merger-driven motions
\citep{Zuhone2011,McNamara2012,Kannan2017,Barnes2019}. It is not
clear how (and if) a suppression of heat transport will modify these
large-scale phenomena.
Such external turbulence (in addition
to the turbulence driven by the MTI itself) could also modify the MTI
evolution by randomizing magnetic field lines
\citep{Ruszkowski2011ApJ,2011MNRAS.413.1295M}.
\citet{2011MNRAS.413.1295M} investigated this and found that the
MTI is able to drive strong turbulence even in the presence of external
turbulence. In particular, they found that the MTI persists on scales larger
than the injection scale of the external turbulence. \citet{2011MNRAS.413.1295M}
did however not include Braginskii viscosity and
suppression of heat conductivity. If the conductivity is suppressed in mirror
unstable regions, then the external turbulence could also modify the MTI
by driving positive pressure anisotropy and creating mirror unstable regions.
Driven turbulence in incompressible, weakly collisional plasmas is under active
investigation \citep{Santos-Lima2014ApJ,Squire2019JPlPh,St-Onge2020}.
While these idealized studies find that a large fraction of the volume is mirror
unstable the mirror unstable volume fraction in turbulence driven
by gas accretion and mergers in galaxy clusters is not well-known.
Alfv\'{e}nic turbulence with Braginskii viscosity driven via an
Ornstein-Uhlenbeck process shows that weakly collisional
systems will have smaller parallel
rate-of-strain,
$\b\b\mathbf{:}\nabla \vec{\varv}$, than found in ideal
MHD \citep{Squire2019JPlPh}. This tendency, termed \emph{magneto-immutability},
means that one cannot simply obtain $\Delta p$ by post-processing a
cosmological simulation performed with ideal MHD and anisotropic conduction
(such as the one by \citealt{Ruszkowski2011ApJ}). While a detailed
understanding of the mirror-unstable fraction in a more realistic,
cosmological setting thus awaits further work, we note for now that we have
found that the saturated state
of the MTI in our setup is rather insensitive to suppression of heat conductivity
over a large range in $\beta_0\Kn$ even when the mirror unstable fraction
is substantial (see Fig.~\ref{fig:parameter_plot}). Taken in combination with
the results of \citet{2011MNRAS.413.1295M}, this indicates
that a study with
external turbulence and $\Sd=0.01$ would find slower growth of the MTI already
at the onset of the simulation but that the nonlinear saturation state would be
very similar to the one found in simulations without turbulence and heat conduction
suppression. Testing this prediction will require a dedicated study taking
both heat conduction suppression and external turbulence into account.

In terms of microphysics, a major limitation of our study is that our subgrid
model for heat conduction considers suppression of heat conductivity by the
ion mirror instability only. Another important candidate for suppression of
heat conductivity is the electron whistler instability
\citep{Komarov2014,Roberg-Clark2016ApJ,Roberg-Clark2018PhRvL,Roberg-Clark2018ApJ,Komarov2018,Riquelme2018}.
This instability could also modify the dynamics and vigorousness of the MTI.
Importantly, while our subgrid model couples the efficiency of heat transport
to plasma motions (via the rate-of-strain tensor and local magnetic field),
suppression by the whistler instability can depend directly on the gradient in
temperature. This occurs because the whistler instability is driven unstable
by the temperature gradient itself \citep{Komarov2018}. Whether the MTI
survives when taking this effect into account is likely to be the topic of
future research (e.g. \citealt{Drake2020arXiv}).
Another complication is that the MTI itself
also exists in the collisionless regime \citep{Xu2016}. Their linear analysis
for a thermally stratified, magnetized and collisionless plasma showed that
the kinetic variant of the MTI also has an electron counterpart. The electron
MTI (eMTI) can grow faster than the standard MTI \citep{Xu2016}. It is however
not clear how the eMTI saturates and whether it modifies subsequent heat
transport. The final conclusion could therefore potentially involve the
outcome of a competition between the eMTI and the electron whistler
instability driven unstable by a thermal gradient.

The goal of our study was to understand whether the MTI is modified by the
mirror instability. In a broader context, however, the motivation was also to
understand whether we can reliably include kinetic effects as subgrid models
in fluid simulations. The MTI here serves as an interesting testbed, as future
PIC simulations could potentially study the collisionless MTI \citep{Xu2016}
in a local 2D setup \citep{Par05}. As global galaxy cluster simulations using
PIC are inconceivable in the foreseeable future, it will inevitably be
necessary to include kinetic effects via subgrid prescriptions. Here we
propose using the MTI, an interesting yet somewhat simple dynamical system
where the microscales drive the macroscales (and vice versa), as a stepping
stone.

\section*{Acknowledgments}

We thank the anonymous referee for constructive reports
with  valuable
comments on plasma physics and the role of external turbulence.
This work was initiated during an extended research stay by TB at the
Theoretical Astrophysics Center (TAC), University of California, Berkeley, and
TB thanks TAC for hosting him. TB and MEP acknowledge support by the European
Research Council under the European Union's Seventh Framework Programme
(FP/2007-2013) under ERC grant agreement 306614. TB and CP
acknowledge support by the European Research Council under ERC-CoG grant
CRAGSMAN-646955. This work was supported in part by a Simons
Investigator Award from the Simons Foundation (EQ). This research was supported in part
 by the National Science
Foundation under Grant No. NSF PHY-1748958.
Our data analysis and plotting was performed with Python
\citep{python}
using Numpy \citep{numpy}, Scipy \citep{scipy} and Matplotlib
\citep{matplotlib}. We thank the developers for making these tools
freely available.

\section*{Data availability}

The data underlying this article will be shared on reasonable request to
the corresponding author. The data was generated with
a modified version of the Athena code, publicly available at
\href{https://princetonuniversity.github.io/Athena-Cversion/}
{https://princetonuniversity.github.io/Athena-Cversion/}.

\bibliographystyle{mnras}
\bibliography{references}

\appendix

\section{Wave disruption by suppression of heat conduction}
\label{sec:soundwave}

\begin{figure*}
\includegraphics[trim= 0 10 0 15]{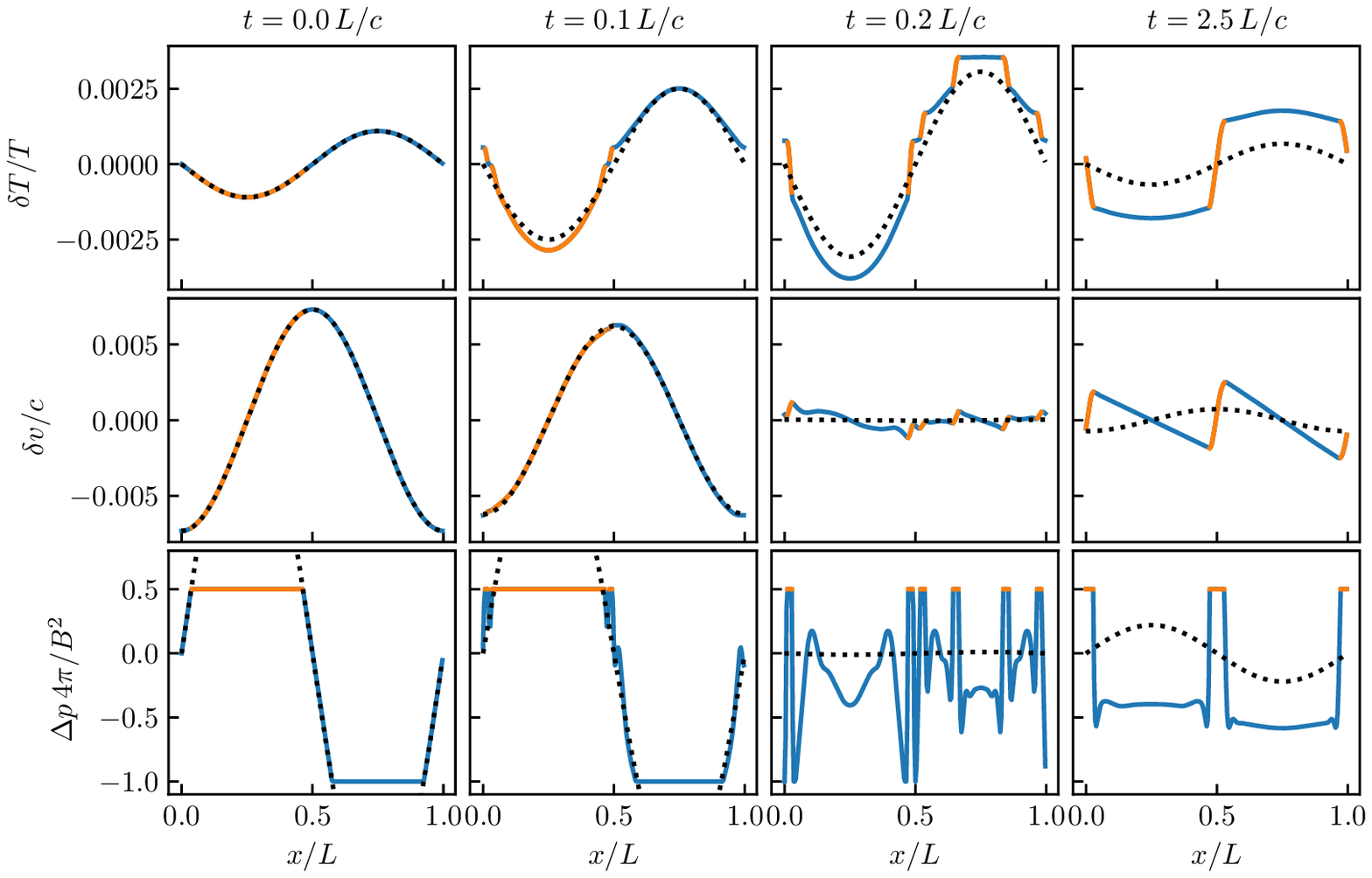}
\caption{Evolution of an acoustic wave in Braginskii-MHD with
a subgrid model for suppression of heat conductivity (which mimicks
reduction by the mirror instability, see Section~\ref{sec:visc-subgrid}).
Dotted lines show small-amplitude solutions to
equation~\eqref{eq:evp}. Solid lines are a numerical solution obtained
with FFT methods. Orange regions have surpassed the mirror threshold,
equation~\eqref{eq:mirror}, and the heat conductivity is reduced by
a factor $S_\mathrm{D}=0.01$ in these regions. The decay of temperature
fluctuations is consequently much slower in these regions and the temperature
distribution becomes terraced (upper row). The pressure anisotropy
(lower row) has a high frequency component which depends on
the grid scale.}
\label{fig:sound_wave_plot}
\end{figure*}

In this appendix, we analyze how sound waves are modified when the heat
conductivity is suppressed in mirror unstable regions. Our
analysis
is a fluid analogue of the kinetic analysis and simulations of
\citet{Kunz2020arXiv}. Sound waves decay via
thermal conduction and viscosity, and their decay rate is expected to change
if the heat conductivity is suppressed. At the same time, sound waves
themselves generate pressure anisotropy (via density changes, see
equation~\ref{eq:pressure_anisotropy}), and sound waves with sufficiently high
amplitude can exceed the mirror instability threshold. The resulting
suppression of thermal conductivity is not uniform across the wave, and
results in a complicated modification to the waveform. Here we illustrate the modified
 wave evolution by using a simple
one-dimensional (1D) setup.

We consider a longitudinal sound wave with small variations in the
temperature, $\delta T$, density, $\delta \rho$, and velocity, $\vec{\varv} =
\delta \varv \ex$. A background magnetic field, $\vec{B} = B \ex$, sets the
direction of anisotropic transport but does not directly enter the dynamics.
The domain is periodic with length $L$ in the $x$-direction. The system can be
modeled with the following set of linearized equations\footnote{
We assume equal ion and electron temperatures
for direct comparison with our study of the MTI. Note, however, that this approximation is expected to break down for sound waves which vary on a fast time scale compared to the ion-electron equilibration time scale (see e.g. \citealt{Zweibel2018,Kempski2020} for a detailed discussion of this).}
\begin{align}
    \pder{}{t} \drho &= - \pder{\delta \varv}{x} \ ,
    \label{eq:lin-rho}
    \\
    \pder{\delta \varv}{t} &= -\f{1}{\rho} \pder{\delta p}{x}
                        +\f{2}{3\rho} \pder{\delta \Delta p}{x} \ , \\
    \pder{}{t}\dt &= - \left(\gamma-1\right)\pder{\delta \varv}{x}
                         + \f{\left(\gamma-1\right)}{p}
                         \pder{}{x}\left(\chi_\para \pder{\delta T}{x} \right)
                         \ ,
\end{align}
where the pressure anisotropy is given by
\be
    \delta \Delta p
    = - 2 \nu_\para \der{\delta \rho}{t} = 2 \rho \nu_\para \pder{\delta \varv}{x}
    \ .
    \label{eq:pa-soundwave}
\en
Here, the effects of anisotropic heat conduction and Braginskii viscosity
are included via the
coefficients $\chi_\para$ and $\nu_\para$. We expect both of these effects to
lead to damping of the wave (due to diffusion of temperature and velocity, respectively).

Fluctuations with sufficiently high amplitude violate the mirror instability
threshold, equation~\eqref{eq:mirror}, and lead to a reduction in the
effective viscosity and heat conductivity.
Building on earlier
work by \citealt{Squire2016,Squire2017,Squire2017b} for Alfv\'{e}n waves,
this effect has very recently been studied for a collisionless system
\citep{Kunz2020arXiv}. For sufficiently small amplitude
fluctuations (such that $\Delta p < B^2/8\upi$ at all times, see detailed
criterion below), solutions can however be adequately described with linear
theory. We find that the system of equations given by
Equations~\eqref{eq:lin-rho}-\eqref{eq:pa-soundwave} yields the eigenvalue
problem
\be
    \left(
    \begin{matrix}
    -\omega & k\cs & 0 \\
    k \cs & -\ui \f{4}{3} \nu_\para k^2 -\omega & k \cs \\
    0 & (\gamma-1) k\cs  & -\ui\f{\gamma-1}{p} \chi_\para T k^2 -\omega
    \end{matrix}
    \right)
    \left(
    \begin{matrix}
    \delta \rho/\rho \\
    \delta \varv /\cs \\
    \delta T/T
    \end{matrix}
    \right)  = 0 \ ,
    \label{eq:evp}
\en
when Fourier transformed in time and space, assuming that $\chi_\para$ and
$\nu_\para$ are constant. Solutions to equation~\eqref{eq:evp} show that
acoustic waves decay at an exponential rate due to both heat conduction and
Braginskii viscosity. Since heat conduction acts on a faster time scale than
viscosity (see equation \ref{eq:kappa_and_nu_para}), heat conduction is
the dominant effect in the decay of such low amplitude waves.

The linear analysis breaks down if the mirror threshold is passed (i.e. if
equation~\ref{eq:mirror} is satisfied). Whether the heat conductivity is
 reduced due to the pressure anisotropy generated by the wave, thus depends on
the amplitude of the sound wave. Using the linear solution for the evolution
of pressure anisotropy, and ignoring the detailed time and space dependence,
the heat conductivity is reduced if
\be
    2 \rho \nu_\para k |A| \, \cs \gtrapprox \f{B^2}{8 \upi} \ ,
\en
where $A \,\cs$ is the initial amplitude of the velocity
perturbation.
In terms of the dimensionless quantities $\beta$, $\mathrm{Kn}^{-1}$ and $kL$,
the limit is therefore predicted to be
\be
    |A| \gtrapprox \f{\mathrm{Kn}^{-1}}{\beta k L} \ .
    \label{eq:int-limit}
\en
Equation~\eqref{eq:int-limit} shows that the sound wave itself most effectively suppress
 the heat conductivity when the magnetic field is weak (i.e. when
$\beta$ is large) or when collisions are rare (i.e. when $\mathrm{Kn}^{-1}$ is
small). Additionally, interruption occurs more easily for short wavelength
waves (i.e. large wavenumbers, $k L$).

We solve equations~\eqref{eq:lin-rho}-\eqref{eq:pa-soundwave}
numerically\footnote{The numerical solutions are found with a fast Fourier
transform (FFT) method for the spatial derivatives and a fourth order
Runge-Kutta time step procedure. We additionally apply a sixth order
hyper-viscosity term to $\delta \rho$, $\delta \varv$ and $\delta T$ to avoid
buildup of grid scale noise.} in order to understand how the wave evolution is
modified when the initial amplitude exceeds the limit given by
equation~\eqref{eq:int-limit}. The simulations are initialized with an
acoustic wave solution obtained\footnote{Three solutions are found: a decaying
solution with zero real frequency, $\omega_0=0$, as well as two waves
traveling to the left and right, respectively. The standing wave solution used
for initializing simulations is constructed from the two traveling waves.}
from the eigenvalue problem given by equation~\eqref{eq:evp}, i.e., $\delta
\varv/\cs = A \cos(k x)$, $\delta \rho /\rho = A_\rho \sin(k x)$, and $\delta
T/T = A_T \sin(k x)$ where $A$, $A_\rho$ and $A_T$ are dimensionless
amplitudes. The linear solution is then a decaying, standing wave with
frequency $\omega = \omega_0 + \ui \Gamma$ where $\omega_0$ is the oscillation
frequency and $\Gamma$ is the decay rate.

We show an example of modified wave evolution in
Fig.~\ref{fig:sound_wave_plot}, with the simulation (solid lines) and the
small-amplitude linear theory for comparison (dashed lines). The simulation
has $kL=2\upi$, $\beta=2\times 10^4$ and $\mathrm{Kn}^{-1}=200$ (here defined
as $\mathrm{Kn} = \lambda_\mathrm{i}/L$). These parameters yield
$\mathrm{Kn}^{-1}/(\beta k L)= 1.6 \times 10^{-3}$ which is less than the
initial velocity amplitude ($|A|=7.3\times 10^{-3}$). This consequently leads
to suppression of heat conductivity by a factor $\Sd=0.01$ in the
mirror-infested regions which are indicated with orange solid lines. Due to
the suppressed conductivity, temperature fluctuations decay at a much slower
rate in the mirror-infested regions than in the regions unaffected by the
mirror instability. This leads to a terraced structure in temperature with
steep slopes in temperature occurring where the conductivity is suppressed and
shallow slopes occurring where the unsuppressed conductivity quickly
erases temperature fluctuations (see in
particular the third panel in Fig.~\ref{fig:sound_wave_plot}).
Note that this modification of the wave structure depends
on the applied suppression factor, $\Sd$. For instance, the modification
is less dramatic when the value motivated by \citet{Kom16}, $\Sd=0.2$,
is used. High-frequency variation in
the heat conductivity is however still found to sporadically occur in sound
wave and MTI simulations with $\Sd=0.2$ (see the online supplementary material).

\begin{figure}
\centering
\includegraphics[trim= 0 25 0 0]{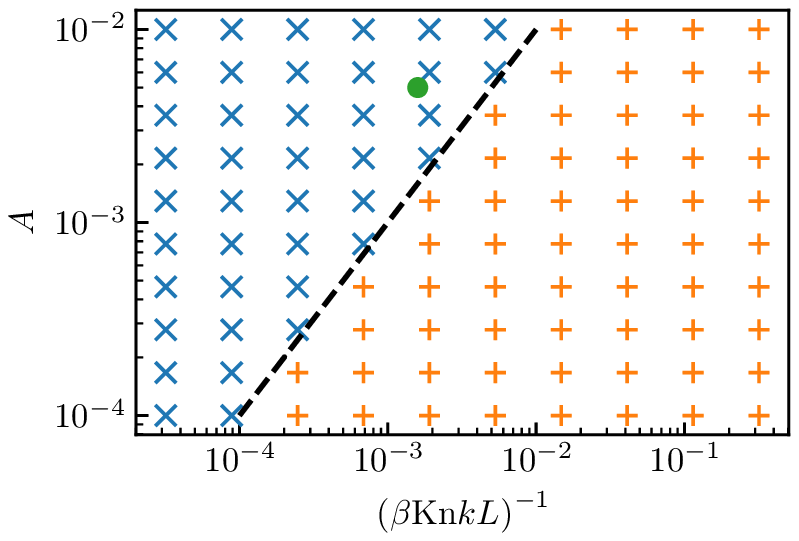}
\caption{Numerical simulations with various initial amplitudes, $A$,
and interruption parameters, $(\beta \mathrm{Kn} k L)^{-1}$.
Data points located above the dashed line, given by equation~\eqref{eq:int-limit},
exceeded the threshold for instability and had suppressed heat conductivity.
The solid green circle corresponds to the parameters that were used for the
simulation presented in Fig.~\ref{fig:sound_wave_plot}.}
\label{fig:mirror_threshold}
\end{figure}

The numerical simulation in Fig.~\ref{fig:sound_wave_plot} shows that
suppression of heat conductivity in mirror-unstable regions can  delay the wave decay.
In Fig.~\ref{fig:mirror_threshold} we compare our analytical prediction for
when this interruption of acoustic waves occurs,
equation~\eqref{eq:int-limit}, with results from a suite of simulations where
we varied the wave amplitude, $A$, and the magnetic field strength, $\beta$.
In this figure, blue crosses indicate simulations in which the
mirror stability threshold was violated and the wave evolution was modified by
mirror instability-induced suppression of the heat conductivity. Below the
amplitude-threshold given by equation~\eqref{eq:int-limit}, orange plus signs
indicate simulations in which the mirror instability threshold was not
violated and the wave evolution decayed in agreement with simple linear
theory.

A limitation of our subgrid model is that sound waves are more likely
to trigger the mirror instability at high wavenumber
(small spatial scales, see
equation~\ref{eq:int-limit}). This causes a grid-scale dependence in the
suppression of heat conductivity (i.e. the striations seen in
Fig.~\ref{fig:mti_evp_imshow}). Despite this grid-scale dependence,
our simulations reach the same result (that the MTI perserveres despite the
suppression of thermal conductivity) at several different levels of numerical
grid
resolution
(see Appendix~\ref{app:resolution-study}). Nevertheless, it would be
beneficial to construct a
regularization method which sets a minimum length scale for regions with
suppressed conductivity. A
well motivated regularization method could perhaps be informed from
collisionless theory or simulations. The mirror instability creates fluctuations in
the magnetic field with a correlation length of $\sim 100-200 \rho_\mathrm{i}$
where $\rho_\mathrm{i}$ is the ion Larmor radius \citep{Kom16}. On smaller
scales, the suppression of heat conduction is no longer efficient, see figure
5 in \citet{Kom16}. This means that our subgrid model for heat conductivity
should be modified at small spatial scales below which the heat conductivity
is not expected to be effectively suppressed.
While these scales are much too small to be resolved in our simulations,
one could artificially increase the cutoff scale in order for it to be resolvable.
Such a regularization method would then remove the
grid-scale dependence of the characteristic size of mirror-infested regions.
Alternatively, one could use an
incompressible code, such as \textsc{snoopy} \citep{Snoopy2015},
to solve the Boussinesq
equations and thereby eliminate sound waves from the
simulation.
However, while the local, linear theory of the MTI can
be derived in the Boussinesq approximation, \citet{2011MNRAS.413.1295M} showed
that local simulations underestimate the saturation amplitude of turbulent
motions compared to a larger box where the vertical extent is not
much smaller than the scale height. Such a quasi-global setup, which we
also use in our work, is at odds with the Boussinesq
approximation which assumes $L_z \ll H$ (e.g. \citealt{Spiegel1960ApJ}).
To get around this
limitation of the Boussinesq approximation, one
could instead employ alternative so-called sound-proof equation sets such
as the anelastic approximation or a pseudo
incompressible model (see \citealt{Brown2012ApJ}, \citealt{Vasil2013ApJ}
and references therein
for a thorough discussion
of these methods for modeling low Mach number, stratified
atmospheres).
In addition, the Mach number, $\mathcal{M} = |\varv|/(\sqrt{\gamma}\cs)$,
in  our non-linear simulations saturate with a
root-mean-square (RMS) value $\mathcal{M}_{\mathrm{rms}} \sim 10^{-2}$
with the largest values
reaching $\mathcal{M}_{\mathrm{max}} \sim 0.1$.
For this reason, modeling the MTI in a local Boussinesq approximation or a
pseudo-incompressible model which assumes $\mathcal{M}\ll1$ would come with
other limitations.
We leave investigating these issues further to future work.

\section{Quasi-global, linear theory for the MTI}
\label{app:quasi-global-theory}

\begin{figure}
\includegraphics[trim= 0 10 0 0]{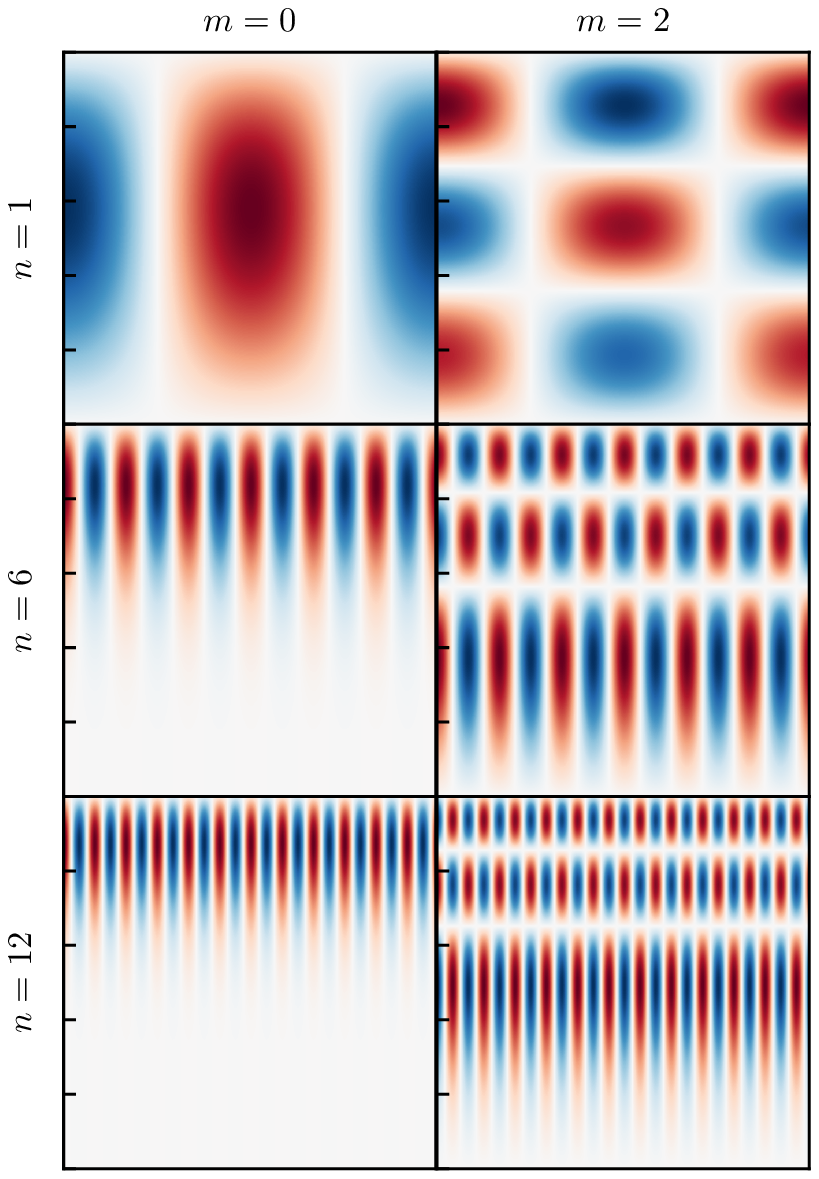}
\caption{The $\dvz$ component of eigenmodes obtained for $k_\para H = 2\upi n$
where $n=1$, 6 and 12 are the horizontal mode numbers and $m=0$ and 2 are the vertical
mode numbers.}
\label{fig:quasi-global-modes}
\end{figure}

Equations~\eqref{eq:rho}--\eqref{eq:energy} are linearized around the
equilibrium given in Section~\ref{sec:numerical_setup}. The perturbations are
assumed to have the form $f(z)\exp(-\ui \omega t + \ui k x)$ where $\sigma = -
\mathrm{Im}(\omega)$ is the growth rate. The key differences between the
present analysis and the one performed in \citet{Kun11} are thus that \emph{i)}
we retain
the $z$-dependence of all background variables instead of assuming locality in
the vertical direction and \emph{ii)} that
we employ reflective boundary conditions at the top and bottom of the domain.
More details on this type of analysis can be found in
e.g. \citet{Lat12,Berlok2016b} where configurations with a vertical magnetic
field leading to the HBI were analyzed.
\begin{figure}
\includegraphics[trim= 0 20 0 0]{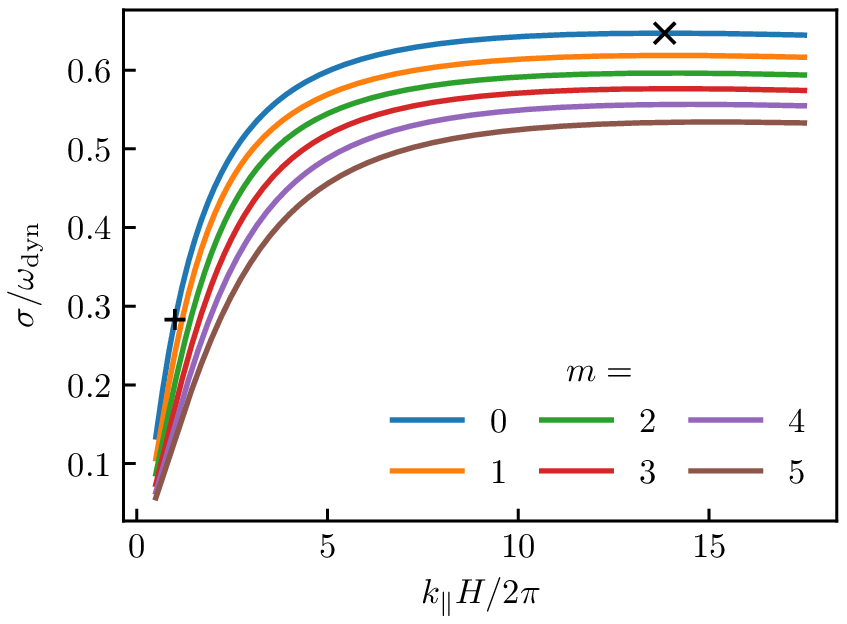}
\caption{Growth rates as a function of the two mode numbers $n=k_\para H/2\upi$ and
$m$ using quasi-global theory. The fastest growing (fundamental) mode is indicated with a $\times$ ($+$) symbol.}
\label{fig:quasi_theoretical_growth_rates}
\end{figure}
The linearized equations for a background horizontal magnetic field are
\begin{align}
  - \ui \omega \drho &=  - \ui k \dvx
    - \left(\der{\ln \rho}{z} + \pder{}{z} \right) \dvz \ ,
    \label{eq:lin-rho-mti}
\end{align}
\begin{align}
  -\ui\omega \dA &= \dvz \ ,
\end{align}
\begin{align}
  -\ui \omega \dvx &= - \ui k \f{\delta p}{\rho} +
  \ui k \va^2 \pder{\ln B}{z} \dA
  - \nu_\para \left(\f{4}{3} k^2 \dvx
   + \ui k \f{2}{3} \pder{\dvz}{z}\right) \ ,
\end{align}
\begin{align}
    -\ui\omega \dvz &= - \f{1}{\rho} \pder{\delta p}{z} -
    \va^2 \left(k^2 + 2\pder{\ln B}{z} \pder{}{z} + \pdder{}{z}\right) \dA
    \nonumber \\
    &-
    \f{1}{\rho}\pder{}{z}\left(\ui k\rho\nu_\para \f{2}{3} \dvx -
    \rho\nu_\para \f{1}{3} \pder{\dvz}{z}
    \right) \ ,
\end{align}
\begin{align}
  -\ui\omega \dt &= - \ui k \f{2}{3} \dvx -
  \left(\der{\ln T}{z} + \f{2}{3} \pder{}{z}\right)\dvz
  \nonumber \\
  & -
   \f{2k^2\chi_\para T}{3p} \left(\dt + \der{\ln T}{z} \dA \right) \ ,
   \label{eq:lin-T-mti}
\end{align}
where $\delta A$ is the perturbed vector potential and the perturbed magnetic
field is obtained by computing $\delta \vec{B} = \nabla \times (\delta A
\ey)$. We solve the eigenvalue problem given by
equations~\eqref{eq:lin-rho-mti}--\eqref{eq:lin-T-mti} with the aid of
\textsc{psecas} \citep{Berlok2019}. In particular, we employ a
Chebyshev-Gauss-Lobatto grid with the boundary condition that $\delta \rho =
\delta \varv_z = \delta A = 0$ at $z=0$ and $z=\HO$. The condition on the
vector potential corresponds to enforcing a horizontal magnetic field at the
boundaries.

\begin{figure}
\includegraphics[trim= 0 25 0 0]{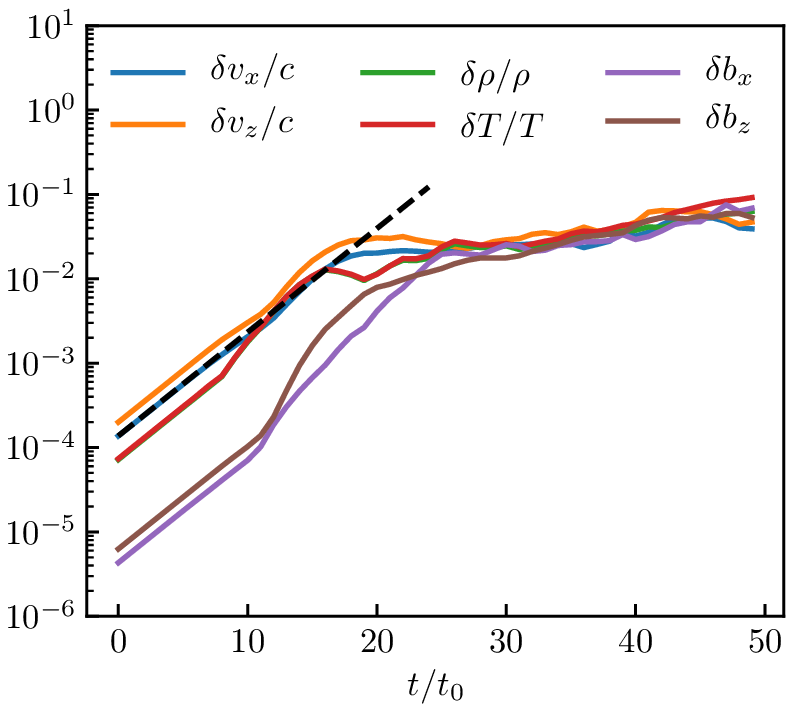}
\caption{Exponential growth of the amplitude of perturbations in
a MTI simulation with $\Sd=0.01$ excited using the fundamental mode $k_x=2\upi/H$
with the fastest growth (i.e. $n=1$, $m=0$). The dashed black line shows the theoretical growth.}
\label{fig:mti-linear-fit}
\end{figure}

We calculate the eigenmodes for an atmosphere with $\beta_0=10^6$ and $\Kni=2000$ as a function of $n = k_\para H/2\upi$.
For each $n$, a set of solutions are obtained which are given an $m$ mode number according to their growth rates (with $m=0$ the fastest, $m=1$ the second fastest, and so on). The $\dvz$ component of a subset of the obtained solutions are shown in
Fig.~\ref{fig:quasi-global-modes}. The growth rates as function of $n$ and $m$ are shown in
Fig.~\ref{fig:quasi_theoretical_growth_rates}.

Due to the reflective boundary conditions, all modes have a vertical variation. It can be observed that
the number of interior zero crossings is given by the $m$ mode number. The modes with high $n$ and low $m$
are stratified with larger amplitude at the top of the domain. As $m$ is increased, the modes penetrate deeper
down into the atmosphere. The growth rates mainly depend on $n$ (see Fig.~\ref{fig:quasi_theoretical_growth_rates}). The variation in growth rate with $m$ is not large and is consistent
with the differences found when using local, linear theory to estimate the growth rate at the top ($m=0$) and bottom (for $m>n$ the modes look sinusoidal in $z$ and penetrate all the way down).

In local, linear theory the fastest growing mode
has $k_z=0$ (see e.g. \citealt{Kun11}).
These are derived in the Boussinesq approximation where $k_x \dvx + k_z \dvz =0$.
The fastest growing mode therefore has $\dvx = 0$ and a constant-with-height
amplitude of $\dvz$. When using reflective boundary conditions, as in our quasi-global theory,
$\dvz$ is enforced to go to zero at the boundaries (see Fig.~\ref{fig:quasi-global-modes}).
As a consequence of this variation with $z$, the modes shown in Fig.~\ref{fig:quasi-global-modes}
have $\dvx\neq0$ and a non-zero velocity divergence, $\nabla \cdot \vec{\varv}\neq 0$.
The pressure anisotropy of a quasi-global eigenmode, given by
$\Delta p = \rho \nu_\para (2 i k_x \dvx - \partial{\dvz}/{\partial z})$,
is consequently linear in the amplitude of the velocity perturbations.
For the fastest growing local mode (which has $k_z=\dvx=0$), the magnitude of the pressure anisotropy is given by
$
    \Delta p = 3 \rho \nu_\para i k_x \dvz \delta b_z
$, i.e., its magnitude is only second order in the perturbation amplitude. For this reason,
the magnitude of the pressure anisotropy in quasi-global modes is larger than for the fastest
growing local mode.

A comparison between the linear growth rate found with
\textsc{psecas} and the simulation presented in Fig.~\ref{fig:mti_evp_imshow}
is presented in Fig.~\ref{fig:mti-linear-fit}. The exponential growth of
perturbations follows the theoretical prediction until mirror-unstable regions
interrupt the mode. Note that the $k H = 2\upi$ mode is not the fastest
growing and that faster growing modes are excited when the interruption
occurs (see Fig.\ref{fig:quasi_theoretical_growth_rates}).

\section{Resolution study}
\label{app:resolution-study}

\begin{figure*}
    \centering
    \includegraphics[trim= 0 10 0 0]{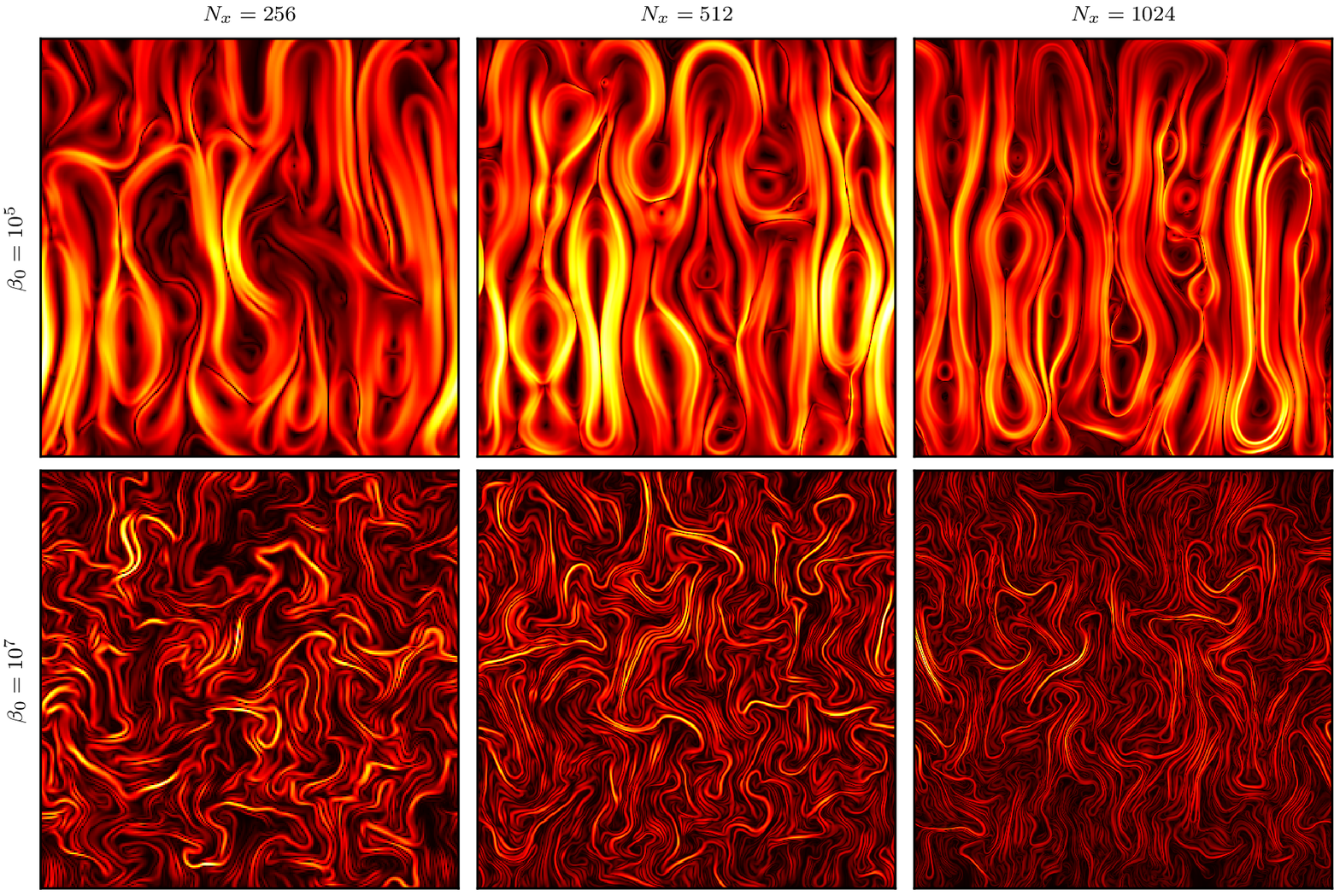}
    \caption{Magnetic field strength, $|B|$, at $t/t_0 = 50$ in 6
    different simulations. A very high spatial resolution is required to
    resolve the thin magnetic field structures that arise in simulations
    in which the magnetic field is weak. The central, unstable part
    of the simulation domain is shown.}
    \label{fig:resolution_images}
\end{figure*}
\begin{figure}
    \centering
    \includegraphics[trim= 0 20 0 0]{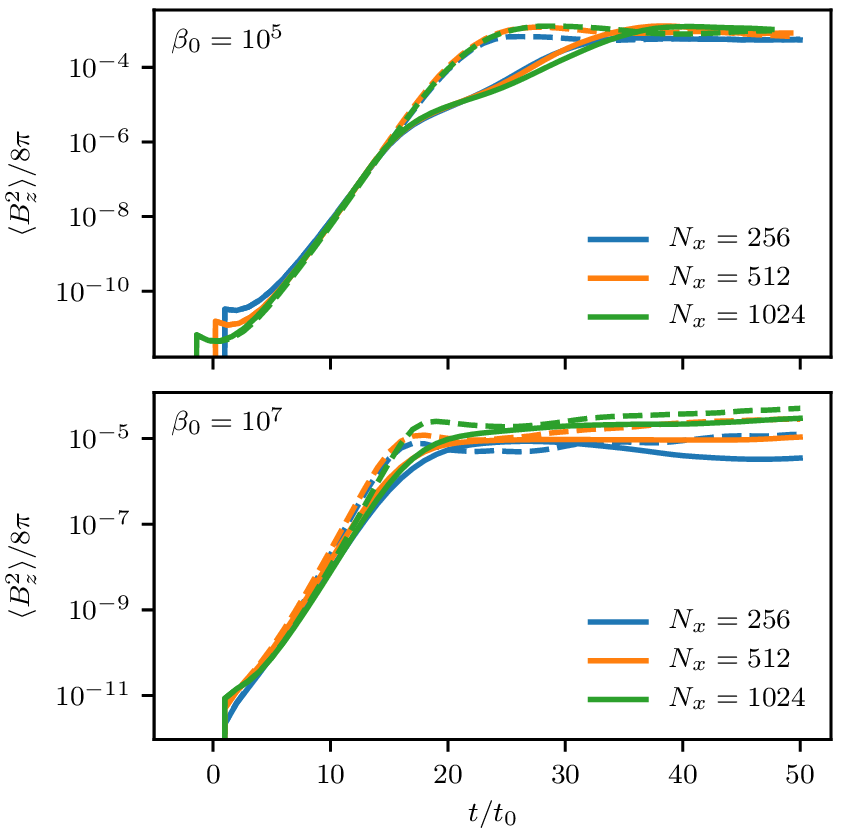}
    \caption{Convergence test of MTI simulations with $\beta=10^5$
    (upper panel) and $\beta=10^7$ (lower panel). Simulations with
    $\Sd=0.01$ ($\Sd=1$) shown with solid (dashed) lines.}
\label{fig:energies_vary_resolution}
\end{figure}

We perform a convergence study of the MTI simulations by considering three
levels of uniform resolution, i.e., $N_x=256$, $512$ and 1024 with twice as
many cells in the $z$-direction.
We show maps of magnetic field strength in
Fig.~\ref{fig:resolution_images} and
magnetic energy evolution in
Fig.~\ref{fig:energies_vary_resolution}.
First, we note that simulations with high $\beta_0$ require
higher spatial resolution in order to resolve thin magnetic field
structures, see Fig.~\ref{fig:resolution_images}.
This is the case regardless of whether $\Sd=0.01$ (as in
Fig.~\ref{fig:resolution_images}) or whether $\Sd=1$.
When performing the parameter study in $\beta_0$ at fixed
$\Kni=2000$ (in
Section~\ref{sec:parameter_study}, see the right-hand column of
Fig.~\ref{fig:parameter_plot}), we therefore
use $N_x=1024$ for $\beta_0=10^7$ and $N_x=512$ for the
$\beta_0=10^6$, $10^5$ and $10^4$ simulations.
Second, we note that decreasing $\Kni$ increases the computational cost of
simulations due to the time step constraint imposed by the parabolic heat
transport term. We are severely limited by the computational cost when
$\Kni=20$, and therefore perform the parameter study where we vary $\Kni$ at
fixed $\beta_0=10^5$ with $N_x=256$ (in Section~\ref{sec:parameter_study}, see
the left-hand column of Fig.~\ref{fig:parameter_plot}). This resolution
appears to be sufficient when $\beta=10^5$, see
Fig.~\ref{fig:energies_vary_resolution}, but would not be sufficient when
$\beta=10^7$.

We consider the evolution of the vertical magnetic energy in simulations
with and without suppression of heat conductivity,
numerical resolution $N_x=256$, $512$ and 1024,
and initial magnetic field strength given by $\beta_0=10^5$ and $10^7$,
see Fig.~\ref{fig:energies_vary_resolution}. In the upper panel where
$\beta_0=10^5$, the evolution of the energy remains
qualitatively and quantitatively the same as the resolution is increased
(with only a very minor systematic increase in the energy of the final state
with resolution). The initial amplitude of the MTI seeding decreases with
resolution (because dissipation of the Gaussian noise by viscosity is more
efficient at high resolution). We have adjusted the starting time
of the $N_x=512$ and 1024 simulations accordingly.
The $\beta_0=10^7$ simulations are not as well converged. Qualitatively,
however, the saturation energy of the $\Sd=0.01$ simulation is always
slightly lower than in the $\Sd=1$ reference simulation.
This can be understood from the sustained presence of mirror-unstable
regions in simulations with $\beta_0=10^7$ and $\Kni=2000$
(see Fig.~\ref{fig:parameter_plot}).

We summarize the resolutions used in the simulations  presented
in the main body of the paper in Table~\ref{tab:simulations}
and the convergence
tests in Table~\ref{tab:convergence-simulations}.

\begin{table}
    \centering
    \caption{Overview of simulations presented in the main body of the
    paper. The left and right columns of Fig.~\ref{fig:parameter_plot}
    are abbreviated as \ref{fig:parameter_plot}L and \ref{fig:parameter_plot}R,
    respectively.}
    \label{tab:simulations}
    \begin{tabular}{lrrll}
        $\beta_0$ & $\Kni$ & Resolution & $\Sd$ & Figure \\
        \hline
        \hline
        $10^6$ & 2000 & $256\times256$ & 0.01 & \ref{fig:mti_evp_imshow}, \ref{fig:mti_evp_profile}, \ref{fig:mti-linear-fit}\\
        \hline
        $10^7$ & 2000 & $1024\times2048$ & 1, 0.01 & \ref{fig:parameter_plot}R\\
        $10^6$ & 2000 & $512\times1024$ & 1, 0.01 & \ref{fig:parameter_plot}R\\
        $10^5$ & 2000 & $512\times1024$ & 1, 0.2, 0.01 & \ref{fig:parameter_plot}R, \ref{fig:energies_vary_factor}\\
        $10^4$ & 2000 & $512\times1024$ & 1, 0.01 & \ref{fig:parameter_plot}R \\
        $10^5$ & 20000 & $256\times512$ & 1, 0.01 & \ref{fig:parameter_plot}L\\
        $10^5$ & 2000 & $256\times512$ & 1, 0.01 & \ref{fig:fiducial_simulation}, \ref{fig:fiducial_lineplot}, \ref{fig:parameter_plot}L\\
        $10^5$ & 200 & $256\times512$ & 1, 0.01 &\ref{fig:parameter_plot}L\\
        $10^5$ & 20 & $256\times512$ & 1, 0.01 & \ref{fig:parameter_plot}L\\
        \hline
        $10^5$ & 2000 & $256\times256\times512$ & 1, 0.01 & \ref{fig:3D-MTI-vis}, \ref{fig:energy_3D_highres}, \ref{fig:power_3D_highres}\\
    \end{tabular}
\end{table}

\begin{table}
    \centering
    \caption{Overview of convergence tests. The lowest resolution was
    $128\times256$ in 2D ($128\times128\times256$ in 3D) and the highest numerical
    resolution performed is stated in the table. All intermediate resolutions
    (i.e. powers of two) were also performed.}
    \label{tab:convergence-simulations}
    \begin{tabular}{lrrl}
        $\beta_0$ & $\Kni$ & Resolution & $\Sd$ \\
        \hline
        \hline
        $10^7$ & 2000 & $1024\times2048$ & 1, 0.01 \\
        $10^6$ & 2000 & $1024\times2048$ & 1 \\
        $10^6$ & 2000 & $512\times1024$ & 0.01 \\
        $10^5$ & 2000 & $1024\times2048$ & 1, 0.01 \\
        $10^4$ & 2000 & $1024\times2048$ & 1 \\
        $10^4$ & 2000 & $512\times1024$ & 0.01 \\
        $10^5$ & 20000 & $256\times512$ & 1, 0.01 \\
        $10^5$ & 200 & $256\times512$ & 1, 0.01 \\
        $10^5$ & 20 & $256\times512$ & 1, 0.01 \\
        \hline
        $10^5$ & 2000 & $256\times256\times512$ & 1, 0.01 \\
    \end{tabular}
\end{table}

\label{lastpage}
\end{document}